\documentclass[fleqn,usenatbib]{mnras}

\usepackage{lsstdesc_macros}

\usepackage{subfig}
\usepackage{graphicx}
\usepackage{threeparttable}
\usepackage{amsmath,amssymb}
\pdfoutput=1

\usepackage{enumerate}
\usepackage{lineno}

\usepackage{hyperref}
\graphicspath{{./}{./figures/}}

\newcommand{\response}[1]{{\textcolor{black}{#1}}}
\newcommand{\responsecwr}[1]{{\textcolor{black}{#1}}}
\newcommand{\responsemnras}[1]{{\textcolor{black}{#1}}}

\usepackage[linesnumbered,boxed]{algorithm2e}

\def\rakurto{\rho^{\text{(4)}}}

\title[PSF Higher Moments Error and WL]{Impact of Point Spread Function Higher Moments Error on Weak Gravitational Lensing}

\author[]{Tianqing Zhang$^1$\thanks{\tt tianqinz@andrew.cmu.edu}, Rachel Mandelbaum$^1$,\newauthor The LSST Dark Energy Science Collaboration.
 \\
$^1$McWilliams Center for Cosmology, Department of Physics, Carnegie Mellon University, 5000 Forbes Ave, Pittsburgh, PA 15213.\\
}

\date{\today}

\begin{document}

\label{firstpage}
\maketitle

\begin{abstract}
Weak gravitational lensing is one of the most powerful tools for cosmology, while subject to challenges in quantifying subtle systematic biases. 
The Point Spread Function (PSF) can cause biases in weak lensing shear inference when the PSF model does not match the true PSF that is convolved with the galaxy light profile. Although the effect of PSF size and shape errors -- i.e., errors in second moments --  is well \responsecwr{studied}, weak lensing systematics associated with errors in higher moments of the PSF model require further investigation.  The goal of our study is to estimate their potential impact for LSST weak lensing analysis. We go beyond second moments of the PSF by using image simulations to relate multiplicative bias in shear to errors in the higher moments of the PSF model. We find that the current level of errors in higher moments of the PSF model in data from the Hyper Suprime-Cam (HSC) survey can induce a 
$\sim 0.05 $ per cent shear bias, making this effect unimportant for ongoing surveys but relevant at the precision of upcoming surveys such as LSST. 

\end{abstract}

\begin{keywords}
  methods: data analysis; gravitational lensing: weak 
\end{keywords}

\section{Introduction}

Gravitational lensing is the deflection of light from distant objects due to the gravitational effects of more nearby objects on the geometry of the Universe. Weak gravitational lensing, or weak lensing, is what occurs in the limit that the lensing deflections are sufficiently weak that they only lead to modest changes in the object's observed shape, size, and flux 
rather than dramatic phenomena such as Einstein rings or multiply-imaged sources. Its sensitivity to the gravitational potential along the line-of-sight makes weak lensing one of the most powerful tools for probing dark matter, dark energy and the growth of large-scale structure of the Universe \citep{Hu:2001fb,2010GReGr..42.2177H,2013PhR...530...87W}. Among all the effects on the galaxy caused by weak lensing, the change in shape, also called shear, is the most accessible signal up-to-date.

The requirements on removal of systematic biases and control of systematic uncertainties in the measurement become tighter as the statistical errors decrease to sub-percent levels, starting with the ongoing or recently completed ``Stage III''  cosmological surveys \citep{2006astro.ph..9591A} such as the Dark Energy Survey \citep[DES;][]{des_review}, 
the Kilo-Degree Survey \citep[KiDS;][]{deJong:2017bkf}, the Hyper Suprime-Cam survey \citep[HSC;][]{2018PASJ...70S...4A}, and the future ``Stage IV'' ground-based surveys such as the Vera C.~Rubin Observatory Legacy Survey of Space and Time \citep[LSST;][]{Ivezic:2008fe, 2009arXiv0912.0201L}, and space-based surveys such as the \textit{Nancy Grace Roman} Space Telescope \citep{2015arXiv150303757S, 2019arXiv190205569A} and \textit{Euclid} \citep{Euclid_overview}.

The Point Spread Function (PSF) is a distribution function that is commonly used to describe the blurring effects of the atmosphere, telescope optics, and pixelization, which convolves the light profiles of the stars and galaxies in the images. 
The PSF therefore changes the observed shape and size of the galaxy. To measure the true galaxy shape despite the convolution with the PSF, one must model the PSF at the galaxy position, \response{based on the images of stars around it}; a variety of methods exist for doing so, e.g., \textsc{PSFEx} \citep{2011ASPC..442..435B} and \textsc{PIFF} (PSF in Full FOV; \citealt{2021MNRAS.501.1282J}).  

Upon obtaining the PSF model at the position of a given galaxy, one can use a variety of methods to measure the shape of the galaxy or its response to weak lensing shear \citep[e.g.,][]{Sheldon:2017szh, Huff:2017qxu}. 
 A mismatch between the PSF model and the true PSF 
 can cause a systematic bias in the weak lensing measurement.  
Previous work in this field has focused on the impact of the errors in the second moments of the PSF model, i.e., differences between the size and shape of the true and model PSF, on the weak lensing shear measurement \citep{Hirata:2003cv, PaulinHenriksson:2007mw, 2010MNRAS.404..350R,2016MNRAS.460.2245J}. Control of residual systematic uncertainties in the shear due to second moment errors in PSF modeling is considered one of the main systematics in weak lensing shear inference 
for both previous surveys such as the HSC survey  
\citep{2018PASJ...70S..25M} and for upcoming surveys such as LSST \citep{2018arXiv180901669T}. 

The formalism derived in the aforementioned papers on this topic would predict zero systematic bias in shear inference as long as the second moments of the PSF model completely match those of the true PSF, neglecting any differences between the higher moments of the PSF model and true PSF. 
\responsemnras{When the PSF is unweighted, the weak lensing shear is only associated with the second moments of the galaxy and PSF. However, a weight function is necessary for shear inference in order to reduce the impact of pixel noise. \cite{2011MNRAS.412.1552M} shows how the lensing shear inference is affected by PSF higher moments when weighted PSFs are used.}
\cite{2020A&A...636A..78S} point out that mismatches between the higher moments of the PSF and the true PSF induce additional multiplicative and additive shear biases on top of those predicted by the second moment formalism in \cite{PaulinHenriksson:2007mw}.  In this paper, we investigate the impact of the higher moments error -- later as HME -- 
on the galaxy shear measurement with image simulations generated using \textsc{GalSim}\footnote{\url{https://github.com/GalSim-developers/GalSim}} \citep{Rowe:2014cza} and real data from the HSC Public Data Release 1  \cite[PDR1;][]{2018PASJ...70S...4A}. 
We simplify the problem by only investigating round PSFs and only considering their radial fourth moments, and the associated multiplicative biases. The goal of this paper is to investigate whether the HME of the PSF model is a significant contributors to systematic biases and uncertainties in the weak lensing shear measurement for LSST. 
 
The structure of this paper is as follows. 
In Section~\ref{sec:background}, we describe relevant background material about weak lensing shear estimation and PSF modeling. We introduce the simulation methods and our analysis of real data in Section~\ref{sec:methods}. In Section~\ref{sec:results},  we show the results of our analysis of the simulations and real data. Based on the results, we derive conclusions about the significance of shear biases caused by the HME of the PSF model in Section~\ref{sec:conclusion}.

\section{Background}
\label{sec:background}


In this section, we summarize background material related to weak lensing shear (Section~\ref{sec:WL_intro}), and the impact of the PSF and PSF modeling  (Section~\ref{sec:PSF_intro}). 

\subsection{Weak Lensing Shear}
\label{sec:WL_intro}

Weak gravitational lensing occurs when light from background objects gets mildly deflected by the intervening matter in the Universe \citep[for a review, see][]{Kilbinger:2014cea}.  
The scientific significance of weak lensing is by no means ``weak'': 
because of its sensitivity to the gravitational potential of the large-scale structure of the Universe, it is a powerful probe of the dark matter distribution and the growth of cosmic structure with time.

Quantitatively, weak lensing is a  
\response{local} linear transformation between the pre-lensing and post-lensing light-ray. The relation between the post-lensing position $(x,y)$ and the pre-lensing position $(x',y')$ can be expressed as 
\begin{equation}\label{eq:sheardef2}
\begin{pmatrix}
x'\\
y'
\end{pmatrix} = (1-\kappa)
\begin{pmatrix}
1-g_1  & -g_2  \\
-g_2 &1+g_1
\end{pmatrix}
\begin{pmatrix}
x\\
y
\end{pmatrix}, 
\end{equation}
where the reduced weak lensing shear $g = g_1 + \mathrm{i}g_2$ is a complex number 
that describes the anisotropic distortion of the galaxies, i.e.\ the shape distortion, and the convergence $\kappa$ is a scalar that describes the isotropic distortion (magnification or contraction) of the observed galaxy. The convergence changes the observed flux and size of the galaxy,  
while the reduced shear changes the shape of the galaxy, e.g., turning round galaxies into elliptical galaxies. 
The first component of the reduced shear, $g_1$, is responsible for the stretch along the x- and y-axes, while the second component, $g_2$, describes the stretch along the diagonal axes at $45^\circ$  
to the x- and y-axes. 
In this paper, we will not consider the convergence effect ($\kappa = 0$), so the shear is the same as the reduced shear.

The mild distortions of galaxy shapes induced by weak lensing, i.e., shear, can only be detected through statistical measurements, often including millions of galaxies \citep{2018PASJ...70S..25M, 2018MNRAS.481.1149Z, 2020arXiv200701845G}. 
Typically the coherent shape distortions induced by weak lensing are measured either via cross-correlation with the positions of galaxies in a massive nearby lens sample \response{(galaxy-galaxy lensing)}, or via auto-correlation of pairs of galaxy shapes \citep[\response{cosmic shear;}][]{2018PhRvD..98d3528T, 2020PASJ...72...16H, 2020arXiv200715633A}. 
The statistical and systematic uncertainties in the shear signal are the main obstacles in making precise cosmological measurements using weak lensing \citep{2013PhR...530...87W}. There are two primary sources of statistical uncertainty in the shear signal \citep{2002A&A...396....1S}: \response{The first one is caused by} the dispersion in the galaxy intrinsic shapes, i.e., shape noise. The second source of statistical uncertainty is due to the large-scale structure that causes various weak lensing signal among the Universe, surveys that observe part of the Universe get a sampling uncertainty, also known as the cosmic variance \citep{2004A&A...413..465K}.

There are multiple source of systematic biases that affect the measurement of the weak lensing shear \citep[for a review, see][]{Mandelbaum:2017jpr}.  
A common approach to systematic biases is to estimate and remove them, either by subtraction from the observed measurement or by modeling the physical processes that generate the biases. Since this correction is in general not perfectly known, even after the correction there will still be some residual systematic uncertainty. Generally, we want the systematic uncertainties to be sub-dominant \response{compared} to the statistical uncertainties. Upcoming  surveys with reduced statistical uncertainties therefore require more stringent control of systematic uncertainty in the weak lensing shear measurement process.

\subsection{PSF Modeling and Systematics}
\label{sec:PSF_intro}

The Point Spread Function (PSF) describes the blurring of astronomical images due to the atmosphere and telescope optics. In practice, we work with the effective PSF, which also 
includes the pixel response function of the detector. The effective PSF convolves the light profiles of the stars and galaxies in the image, which changes the observed size and shape of the stars and galaxies. 

Inferring the weak lensing shear distortion using information about the pre-PSF galaxy shape given the convolved image and PSF model is a substantial challenge. In the GREAT3 challenge \citep{2015MNRAS.450.2963M}, numerous shape measurement methods are tested and compared using simulations. Some of the methods shown there have been used in weak lensing survey science since then, e.g., re-Gaussianization in HSC \citep{2018PASJ...70S..25M}, metacalibration and im3shape in DES \citep{2018MNRAS.481.1149Z}, and shear calibration using pixel-level simulation \response{(lensfit) in CFHTLenS and KiDS \citep{2013MNRAS.429.2858M, 2017MNRAS.467.1627F, 2019A&A...624A..92K}}. However, several principled shear inference methods have been developed \response{which} should work to very high precision by avoiding the sources of bias in earlier methods \citep{Sheldon:2017szh, 2020ApJ...902..138S, 2016MNRAS.459.4467B} 
at least for isolated galaxies -- but they do rely on an accurate PSF model.

There are two main categories of PSF modeling methods:  empirical approaches that rely on the data in the images, and analytical approaches that simulate the physical processes of the PSF \citep{Mandelbaum:2017jpr}. The analytical approach is more commonly used in analysis of the space-based telescopes \response{due to deterministic light propagation}, e.g., the \textit{Hubble} Space Telescope (HST) and its COSMOS weak lensing analysis (\citealt{2007ApJS..172..219L}; see also \citealt{2020MNRAS.496.5017G} for methods to assess model fidelity). Analysis of data from ground-based telescopes has tended to utilize empirical PSF models \response{due to the stochastic nature of the atmosphere}, e.g., the DES Y1 catalog \citep{2018MNRAS.481.1149Z}, the KiDS-1000 catalog \citep{2020arXiv200701845G} and the first-year HSC catalog \citep{2018PASJ...70S..25M}. \responsecwr{The PSF profiles for space-based and ground-based telescopes are usually very different, because of the existence of the atmospheric PSF for the ground-based telescopes. For the purpose of this paper, we focus on the ground-based telescope PSF. } The first step is to measure the effective PSF from a set of stars (typically isolated and with high signal-to-noise ratio) in the image -- we refer to these as the PSF stars. Then the PSF at other positions is obtained by interpolation of the PSF model inferred from the PSF stars.  Out-of-focus wavefront sensing has been \response{recently} \responsecwr{developed} to model the optical PSF \citep[e.g.,][]{1995ApOpt..34.4951K, 1982OptEn..21..829G, 10.1117/12.2056904, 10.1117/12.2234456} \response{and} \responsecwr{a composite PSF model with wavefront modeling of the optical PSF component is planned to be used in future DES releases \cite{2021MNRAS.501.1282J}.} The coadded image is a combination of several images at a given point on the sky,  which has implications for its PSF.  For example, in the HSC pipeline, the coadded PSF is generated in a principled way through weighted averaging of individual exposures, resulting in a well-defined PSF model also based on weighted averages  \citep{2018PASJ...70S...5B}.

The limited information on the spatial and temporal variation of the PSF for ground-based telescopes leads to some intrinsic limitations in the PSF model fidelity. Moreover, errors in modeling some detector effects, such as the brighter-fatter effect \citep{2014JInst...9C3048A} and the \responsemnras{interpixel capacitance of the complementary metal-oxide-semiconductor (CMOS) detectors \citep{old..interpixel,2016PASP..128i5001K}}, can also drive errors in PSF models. 
Most of the commonly-used tests to determine the quality of PSF modeling rely on estimates of PSF and star sizes and shapes, mathematically defined using the observed second moments of the images, e.g., most tests in \cite{2018PASJ...70S...5B}. 
The weighted second moment $Q_{ij}$ of a light intensity profile $f(x)$ 
is defined as 
\begin{equation}
\label{eq:secondmoment}
    Q_{ij} = \frac{1}{F^{(0)}} \int \mathrm{d}x_i \mathrm{d}x_j (x_i - x_i^{\text{cen}})(x_j - x_j^{\text{cen}}) f(\boldsymbol{x}) \omega(\boldsymbol{x}).
\end{equation}
\responsemnras{Here $\omega(\boldsymbol{x})$ is the adaptive Gaussian weight that has a size and shape matched to that of the light intensity profile, centred at the centroid of the profile \citep{Hirata:2003cv}. The weight is introduced to reduce the effect of noise in real images; however, it is the reason that the PSF higher-moments affect shear measurement \citep{2011MNRAS.412.1552M}. }  $F^{(0)}$,  the normalization factor, is the total flux of the light profile weighted by $\omega(\boldsymbol{x})$.  
$x_i^{\text{cen}}$ is the weighted centroid in the $i^{th}$ dimension, calculated as
\begin{equation}
x_i^{\text{cen}} = \frac{1}{F^{\text{(0)}}} \int \mathrm{d}x_i \mathrm{d}x_j x_i f(\boldsymbol{x}) \omega(\boldsymbol{x}).
\end{equation}
    
The weighted second moments radius of the light profile can be defined as $\sigma = \sqrt{(Q_{11}+Q_{22})/2} = \sqrt{T/2}$, where $T$ is the trace of the second moment matrix.  The ellipticity can be defined as $e_1 = (Q_{11}-Q_{22})/(Q_{11}+Q_{22}) =(Q_{11}-Q_{22})/T $ and $e_2 = 2Q_{12}/(Q_{11}+Q_{22}) = 2Q_{12}/T$. The $e_1$ and $e_2$ are related to the axis ratio and position angle of the galaxy ellipse. Like the reduced shear $g = g_1 + i g_2$, the shape is also a spin-2 quantity. Therefore, for the rest of the paper, we denote shape as $e = e_1 + i e_2$ and the amplitude of the shape as $|e| = \sqrt{e^*e}$. 

\cite{PaulinHenriksson:2007mw} explored the systematic biases in weak lensing shear measurement associated with errors in modeling the second moments of the PSF. The bias in the measured ellipticity of the galaxy 
$\delta e^{\text{sys}}$ is
\begin{equation}\label{eq:desys}
    \delta e^{\text{sys}}\simeq (e_{\text{gal}} - {e}_{\text{PSF}})\frac{\delta (R_{\text{PSF}}^2)}{R^2_{\text{gal}}} - \left( \frac{R_{\text{PSF}}}{R_{\text{gal}}}\right) ^2\delta {e}_{\text{PSF}} ,
\end{equation}
where $R_{\text{gal}}$ and $R_{\text{PSF}}$ are the radius of the pre-PSF galaxy and the PSF, respectively.
The two error terms, $\delta(R^2_{\text{PSF}})$  and $\delta {e}_{\text{PSF}}$, are typically referred to as the PSF size and shape error, respectively. Eq.~\eqref{eq:desys}  
is usually used to place requirements on the quality of the PSF model, given some requirement on the control of systematic biases in the weak lensing shear  $\delta{e}^{\text{sys}}_{\text{PSF}}$. Based on the formalism above, requirements can be placed on tolerance for systematic uncertainty in second moments of the PSF model for weak lensing \citep[e.g., Section~3 of][]{2018PASJ...70S..25M}.  \response{This formalism is exactly correct when the weight $\omega(\boldsymbol{x}) \equiv 1$ or when both the galaxy and the PSF are Gaussian with a Gaussian weight. In a realistic scenario, neither of these conditions will be met. Therefore, this formalism cannot be used to predict precise numerical values for the shear biases caused by PSF second moment errors, though it still provides an approximate estimate of their magnitude and trends with galaxy size. }

The formalism described above for estimating weak lensing systematic biases and uncertainties induced by PSF modeling errors only considers the PSF second moments, not any of the higher moments of the PSF model. 

\response{In this paper, we explore shear biases directly associated with the higher moments modeling error of the PSF, by conducting image simulations with deliberated added HME to the PSF model, and by comparing real PSF model images to star images. In our approach, we focus on PSFs in ground-based observations.  We will show that the HME of the PSF modeling (at least given current PSF modeling algorithms) contributes non-negligible systematic error for Stage IV ground-based weak lensing surveys. }

In this paper, we demonstrate methodology and project shear biases due to higher moment errors of the PSF using a PSF modeling method called \textsc{PSFEx} 
\citep{2011ASPC..442..435B}.  This method has been used in practice for weak lensing science in HSC \citep{2018PASJ...70S...5B,2018PASJ...70S..25M} and DES Y1 \citep{2018PhRvD..98d3528T}. Although Rubin's LSST science pipelines are unlikely to use \textsc{PSFEx} for LSST itself, assessing the status of algorithms that are currently in use can help us understand the current level of PSF modeling error and its impact on weak lensing science, and place requirements on future performance.

\section{Methods}
\label{sec:methods}

In this section, we describe key analysis methods used for this work. The first approach we take to quantifying the relationship between weak lensing shear systematics and the HME of the PSF model uses image simulations. Before diving in the simulation step, we first define the quantities we measure for the higher moments in Section~\ref{sec:repofhome}, and explore the shear measurement methods that we are taking in Section~\ref{sec:shapemeasurement}. In Section~\ref{sec:simulation}, we introduce the inputs and steps for producing the image simulations. Section~\ref{sec:hscdata} describes the approach to inspecting the PSF and its model in real data from the HSC survey. While the simulations enable us to relate the HME of the PSF model to a shear bias, the real data provides an estimate for the current level of HME in PSF models in real data. 

\subsection{Higher Moments}
\label{sec:repofhome}

In this subsection, we introduce how the moments of light profiles are defined and computed in practice.

In principle, carrying out our study requires a method for measuring any higher moment of PSF light profiles (beyond second moments), and for introducing a controlled variation in individual higher moments while preserving the second moments.  However, for this initial pilot study we consider a simplification, and quantify the impact of deviations only in the weighted radial fourth moment (defined below).  In practice, we recognize that other higher moments may be relevant, but we defer a detailed decomposition to future work, focusing here on a rough order-of-magnitude estimate of the importance of the higher moments of the PSF for weak lensing.

In practice, we measure 
the \responsemnras{standardized} weighted radial $4^{\text{th}}$ moment, or kurtosis $\rho^{(4)}$, using 
\textsc{GalSim} \citep{Rowe:2014cza}. 
For a light profile $f(\mathbf{x})$, this quantity is defined as
\begin{equation}
\label{eq:kurtosis}
\rho^{(4)} = \frac{ \int (r/\sigma)^4 f(\mathbf{x}) \omega(\mathbf{x}) \mathrm{d}\mathbf{x} }{\int f(x )\omega(\mathbf{x})\mathrm{d}\mathbf{x}}. 
\end{equation}
where $r = |\mathbf{x}|$ 
and $\omega(\mathbf{x})$ is the adaptive Gaussian weight we used in Eq.~\eqref{eq:secondmoment}, $\sigma$ is the second moment radius. \responsemnras{The superscript of $\rho^{(4)}$ is a notation for the kurtosis, rather than the $4^{\text{th}}$ power.} The denominator is a normalization factor.  The weighted radial kurtosis for some common PSF profiles is listed in Table~\ref{tab:kurtosis}. 
\responsemnras{The Airy PSF has an undefined second moment $\sigma$ and kurtosis when calculated without a weight function, and the weighted moments depend strongly on the choice of weight function.  The kurtosis value we show in Table~\ref{tab:kurtosis} is calculated with an adaptive Gaussian weight function with $\sigma_{\text{w}} = 0.41 \lambda/D$, where $D$ is the diameter of the aperture. This algorithm-generated weight function has the size proportional to the PSF size, and keeps the Airy profile well-sampled for moment measurements.   }
For the rest of the paper, we define the fractional kurtosis bias $B[\rakurto]$ as 
\begin{equation}
\label{eq:kurtosis_error}
B[\rakurto] = \frac{(\hat{\rho}^{(4)} - \rho^{(4)})}{\rho^{(4)}}
\end{equation}
where $\hat{\rho}^{(4)}$ is the model kurtosis and $\rho^{(4)}$ is the true kurtosis.

\begin{table}
\begin{center}
\begin{tabular}{ccc}
\hline
 Profile       & $\rho^{(4)}$   \\\hline
 Gaussian      & 2.00           \\
 Kolmogorov    & 2.09           \\
 Moffat, $\beta = 3.5 $        & 2.11           \\
 S\'ersic, n = 1 & 2.35                \\
 S\'ersic, n = 4 & 2.74         \\
 Airy PSF, $\lambda = 750 \text{nm}$, $D = 8 \text{m}$ & 1.91\\\hline     
\end{tabular}
\end{center}
\caption{The weighted radial kurtosis value $\rho^{(4)}$ for commonly-used light intensity profiles. The kurtosis is measured using images with gradually decreasing pixel scale to the point that the kurtosis value converges to the second decimal place.  
Note that these are the radial kurtosis values for the named profiles themselves, without any additional pixel response function. \responsemnras{The Airy profile is that for an 8-meter aperture telescope at $\lambda = 750$~nm. The kurtosis is calculated with a Gaussian weight function with $\sigma_{\text{w}} = 0.41 \lambda/D$, proportional to the scale of the Airy profile. }
}
\label{tab:kurtosis}
\end{table}

Again, to reiterate, while we quantify the impact of higher moments error (HME) of the PSF using the radial weighted kurtosis, in general not only the kurtosis but rather all higher \response{radial} moments are perturbed. \responsemnras{All higher moments referred to throughout this paper are the scale-independent standardized weighted moments. }

\subsection{Shape and Shear Measurement}
\label{sec:shapemeasurement}

When measuring cosmological weak lensing, there are methods that measure the shape of individual galaxy and then take the average shape to measure shear. There are also methods that directly act on galaxy ensembles to measure shear. 
For the first category, the shape measurement method is a crucial element in the pipeline. For this reason, we investigate the bias on the outcome of shape measurement, for single galaxies, as a step toward understanding the impact of HME on weak lensing shear. It is important to notice that the shape biases we investigate are not induced by intrinsic limitations of the shape measurement methods, and we are expecting to get different responses from different methods. 
One commonly used shape measurement method is the re-Gaussianization \citep{Hirata:2003cv} method implemented in the HSM module \citep{Mandelbaum:2005wv} in \textsc{GalSim} \citep{Rowe:2014cza}. To test for different responses to errors in the higher moments of the PSF model, we also carried out limited testing with the linear \citep{Hirata:2003cv, Bernstein:2001nz} and KSB \citep{Kaiser:1994jb} methods as implemented in \textsc{GalSim}.

To ensure that our results reflect galaxy shape or ensemble shear biases due to errors in the higher moments of the PSF model, rather than reflecting limitations in the shape measurement methods, we perform each measurement twice. The first measurement $e$ or $g$ uses the true effective PSF, and the second measurement $\hat{e}$ or $\hat{g}$ uses the model effective PSF. The difference between the two measurements, $\hat{e} - e$ or $\hat{g} - g$, is the shape or shear bias we are interested in, denoted as $\delta e$ or $\delta g$. 

\response{In real weak lensing observations, very large galaxy ensembles are typically measured to beat down the intrinsic shape noise. 
However, in the image simulations, we can bypass this problem using the approach from \citet{Massey:2006ha} of creating a 90-degree rotated counterpart for each galaxy before applying the cosmological lensing shear.   
We refer to a galaxy and its rotated counterpart as a 90-degree rotated pair; the galaxies in each pair have opposite values of $e_1$ and $e_2$ in the absence of lensing shear, so (especially in simulations without pixel noise added) a very small number of galaxy pairs can be used to efficiently assess the level of ensemble shear estimation bias.
The shear bias of the 90-degree rotated pair} is calculated by 
\begin{equation}
\label{eq:define_shear_bias}
    \delta g = \frac{(\hat{e}+\hat{e}_{90}) - (e+e_{90})}{2},
\end{equation}
where $\hat{e}$ and $\hat{e}_{90}$ are the shape of the original and the rotated galaxy, measured using the model PSF, and $e$ and $e_{90}$ are measured using the true PSF. When we have more than one galaxy and its pair, which we will call a galaxy ensemble, the ensemble shear is the average over different galaxy shear $\langle g \rangle$.
The ensemble shears $\langle g \rangle$ and shear biases $\delta \langle g \rangle$ are estimated in an analogous process, with $\hat{e}$ replaced by the average shape $\langle \hat{e} \rangle$ in Eq.~\eqref{eq:define_shear_bias}.

\response{While significantly increasing the efficiency of the simulation and decreasing the statistical uncertainty on the ensemble shear, this approach using 90-degree rotated pairs has its limitations.  For example, it limits our ability to measure selection bias; however, this is not the focus of this paper.}

In addition to these older battle-tested methods whose limitations are well-understood, we also use metacalibration \citep{Sheldon:2017szh,Huff:2017qxu}, a state-of-the-art method that self-calibrates multiplicative and additive bias in the ensemble shear inference. The goal of doing so is to check how sensitive our results are to the choice of shear inference method. We use the implementation of metacalibration in the publicly-available  \textsc{ngmix}\footnote{\url{https://github.com/esheldon/ngmix}} package.

\subsection{Image Simulation}
\label{sec:simulation}

Here we describe the image simulation procedure used in this paper. The objects we simulate are postage stamp images of PSF-convolved galaxies and PSFs.  
For each step, we first explain the general settings for all simulations, and then provide details of different simulations. The parameters used in some of the  simulations are tabulated in Table~\ref{tab:specification}. 

For all of the image simulations, we will need to generate two types of postage stamp images with \textsc{GalSim} objects: the observed image of the isolated galaxy convolved with true PSF, and the image of the PSF, with or without kurtosis error. All images are generated with a pixel scale of $0.2 \arcsec$, similar to the pixel scale of the LSST camera. The images are rendered using the Fourier Transform method in \textsc{GalSim}, and include the pixel response function. The PSF-convolved galaxy and true PSF images are then used to estimate the single galaxy shape $e$, 90-degree rotated pair shear $g$ and the ensemble shear $\langle g \rangle$.  The PSF-convolved galaxy and model PSF images are used to estimate the single galaxy shape $\hat{e}$, 90-degree rotated pair shear $\hat{g}$ and the ensemble shear $\langle \hat{g} \rangle$. No noise is included in the images. 

The galaxy profiles that we simulate as specified in Section~\ref{sec:galaxyprofile}, and the  PSF profiles specified in Section~\ref{sec:PSF_profile}, exhibit a gradual increase in complexity and realism. We provide a general roadmap to our simulations in Section~\ref{sec:simulation_roadmap}.  There we describe how the simpler simulations help us develop intuition about the main parameters that determine shear biases for a given PSF kurtosis bias, while the complicated simulations provide a more realistic estimate of ensemble shear biases due to PSF kurtosis bias for LSST.

\begin{table*}
\begin{tabular}{cccccc}
\hline
Index &Galaxy Type & Galaxy Parameters                                                               & PSF Type   & PSF Parameters                   & Fractional Kurtosis Bias $B[\rakurto]$  \\ \hline
1 & Gaussian    & $\text{FWHM}$ = 1.2$\arcsec$                                                                   & Gaussian   & $\text{FWHM}$ = 0.7$\arcsec$, 1.2$\arcsec$, 1.6$\arcsec$ & -0.01 -- 0.01                 \\
2 & Gaussian    & $\text{FWHM}$ = 2.4$\arcsec$, 12.0$\arcsec$                                                               & Gaussian   & $\text{FWHM}$ = 0.9 $\arcsec$ -- 4.7$\arcsec$        & $\sim 0.005$           \\
3 & Gaussian    & $\text{FWHM}$ = 2.35$\arcsec$, 12.0$\arcsec$                                                               & Kolmogorov & $\text{FWHM}$ = 0.7$\arcsec$ -- 3.3$\arcsec$          & $\sim 0.005$             \\
4 & Gaussian    & $\text{FWHM}$ = 2.35$\arcsec$, 12.0$\arcsec$                                                               & Moffat & $\text{FWHM}$ = 0.7$\arcsec$ -- 3.3$\arcsec$          & $\sim 0.005$             \\
5 & S\'ersic      & \begin{tabular}{@{}l@{}} $R_h$ = 1$\arcsec$, 5$\arcsec$ \\  $n =$ 0.5, 1.5, 3\end{tabular}  & Gaussian   & $\text{FWHM}$ = 0.9$\arcsec$ -- 4.7$\arcsec$        & $\sim 0.005$ \\
6 & S\'ersic      & \begin{tabular}{@{}l@{}} $R_h$ = 1$\arcsec$, 5$\arcsec$\\  $n = 0.5$, 1.5, 3\end{tabular} & Kolmogorov & $\text{FWHM}$ = 0.7$\arcsec$ -- 3.3$\arcsec$          & $\sim 0.005$  \\       \hline       
\end{tabular}
\caption{\label{tab:specification}
The specification of the galaxy, PSF and shape measurement methods in the single galaxy simulations described in Section~\ref{sec:simulation}. The commas in the table denote a list of values for which simulations were carried out. 
The ``--'' means we make simulations covering a range of values between the two endpoints shown.  
The kurtosis error is given by Eq.~\eqref{eq:kurtosis_error}. 
The half light radius, or ``$R_h$" in the table, is used to define the size of the S\'ersic galaxies. \response{The kurtosis bias $B[\rakurto]$ changes by a few percent when constructing the model PSF and rescaling to ensure the PSF second moment is preserved; in our analysis we use the actual measured $B[\rakurto]$ rather than the idealized value in this table.}
} 
\end{table*}

\subsubsection{Galaxy Profile}
\label{sec:galaxyprofile}

The first step of the image simulation process is to define the galaxy profiles that we are going to simulate. Our approach is to generate simulations that include galaxy profiles with increasing complexity. First we generate a single (non-round) Gaussian galaxy, then we add complexity to the model to include S\'ersic profiles. We generate 90-degree rotated galaxy pairs, as described in Sec.~\ref{sec:shapemeasurement}, to certify the results we get from single galaxy experiments.  Finally, we generate ensembles of galaxies that include a range of profiles, sizes, and shapes, similar to that found in real data. \response{We also generate 90-degree rotated pairs for the galaxies from the catalog, to eliminate the shape noise.} To gain intuition, we start by quantifying shape measurement biases for  single galaxy experiments. We later proceed to study ensemble shear biases.

The first and simplest galaxy profile we generate is a 2d Gaussian profile, specified by its size $\sigma$ and ellipticity $e$. 
We alter the parameters of the Gaussian galaxy to see its impact on the kurtosis induced shape/shear bias. The size of the Gaussian galaxies range from $\text{FWHM}=1.2''$ to $12.0 '' $, and the ellipticity of the Gaussian galaxies are altered between $|e| = 0.0$ to $0.5$. 

A more realistic and commonly-used 
\citep{1997AJ....113..531D} galaxy model is the S\'ersic 
profile \citep{1963BAAA....6...41S}. A round S\'ersic profile $f_{\text{S\'ersic}} (R)$, is given by
\begin{equation}
    f_{\text{S\'ersic}} (R) = \exp{[-b_n (\frac{R}{R_h})^{\frac{1}{n}}]},
\end{equation}
where $R_h$ is the half light radius \responsemnras{of the single S\'ersic profile}, and $n$ is the S\'ersic index. $b_n$ is a scaling factor to make sure the profile has the correct half light radius; its value is pre-determined for a fixed S\'ersic $n$.  
We first carry out experiments that simulate individual S\'ersic galaxies with a chosen value of S\'ersic index and $R_h$, along with their 90-degree rotated pair, to examine the relation between shear bias and S\'ersic profile parameters.  After that point, we proceed to simulate galaxy ensembles with realistic size, shape, and S\'ersic index distributions.

\begin{figure*}
    \centering
    \includegraphics[width=2.00\columnwidth]{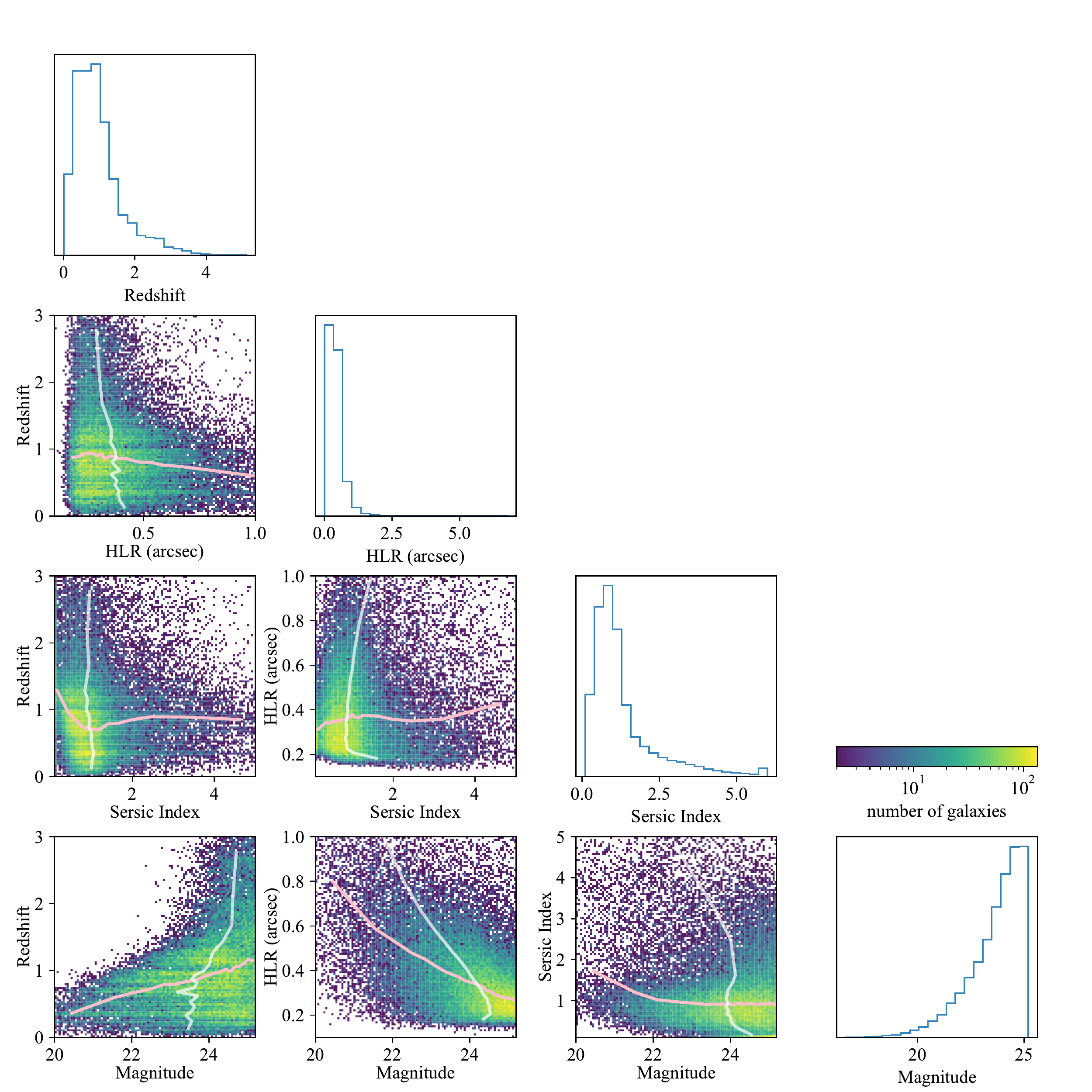}
    \caption{One- and two-dimensional histograms of galaxy properties including redshift, \responsemnras{half light radius},  F814W magnitude and S\'ersic index in the COSMOS parametric catalog, with the pink trend-lines on each panel showing the median of the properties on the vertical axis when binned by the properties on the horizontal axis, \response{and white trend-lines showing the median of the horizontal properties binned by the vertical properties. One-dimensional histograms of individual properties are shown on top.} The color of the plot represents the number of galaxies on a logarithmic scale. \responsemnras{The single S\'ersic profile fitted to the COSMOS galaxy is parameterized by the half light radius and the S\'ersic index shown.} 
    The relationships between parameters can be used to explain the redshift-dependent shear biases. 
    }
    \label{fig:great3_properties}
\end{figure*}

To study the shear biases of galaxies with a realistic distribution of sizes, shapes and S\'ersic indices,   
we use a sample of galaxies from COSMOS with S\'ersic fits\footnote{\url{https://github.com/GalSim-developers/GalSim/wiki/RealGalaxy-Data}} \citep{2015MNRAS.450.2963M}, for which \textsc{GalSim} has a class defined so as to efficiently use the sample for image simulations.  
The COSMOS parametric galaxy catalog that we use includes S\'ersic profile fits to the real galaxy images in the COSMOS HST survey \citep{2007ApJS..172..196K} 
for galaxies down to a limiting magnitude of F814W$=25.3$. Without any cut on the galaxy population, we have $\sim 50,000$ galaxies and its 90-degree rotated-pair to work with.  We use the fits to single S\'ersic profiles with the S\'ersic index allowed to vary 
to generate the S\'ersic galaxy samples using the \code{COSMOSCatalog} module in \textsc{GalSim}. \response{The centroids of the generated galaxies are randomly displaced (by a uniform distribution) within one pixel from the centre of the images. We have confirmed that this displacement does not affect the overall results.} To generate a galaxy population similar to what LSST may use for weak lensing shear inference, we impose a cut on the resolution factor $R_2$ defined by
\begin{equation}
\label{eq:resolution_factor}
    R_2 = 1 - \frac{T_P}{T_I},
\end{equation}
where $T_P$ is the second moment trace of the PSF and $T_I$ is the trace of the PSF-convolved image. The galaxy is well-resolved when $R_2 \sim 1$ and poorly-resolved when $R_2 \sim 0$. 
As suggested by \cite{2018PASJ...70S..25M}, we use galaxies with $R_2 > 0.3$. After the cut, we are left with $\sim 41,000$ galaxies and their 90-degree rotated-pairs.
By drawing randomly from this catalog, we hope to mimic the observed galaxy light profiles for a sample of galaxies such as would be used for an LSST cosmology analysis.  While S\'ersic profiles do not include some of the more complex features of realistic light profiles, we will observe that the simulations with Gaussian versus more general S\'ersic light profiles do not exhibit very different behavior with respect to shear biases due to errors in the higher moments of the PSF.  For that reason, we consider the omission of more complex light profiles to be acceptable in this pathfinder study. After we create the 90-degree rotated pairs, we apply the same amount of shear to these galaxies as the signal we are measuring.

\response{Since there is statistical uncertainty due to cosmic variance, the COSMOS galaxies are not fully representative of the full distribution of galaxy properties \citep{2015MNRAS.449.3597K}.  This is particularly an issue when binning the galaxies by redshift, so that an even smaller volume is being sampled than when using the entire COSMOS volume.  However, even if they were a representative sample, we would still need to determine how many galaxies we must sample from the COSMOS parametric catalog so as to reduce the statistical uncertainty due to the limited number of samples necessary level.}  We do this based on the statistical uncertainty in the shear bias measured using subsamples of galaxies from the catalog and their 90-degree rotated pairs in the absence of pixel noise. Since the systematic shear biases that we are interested in constraining are at the 0.1\% level, we want shear biases to be measured at least one order of magnitude more precisely than that. We determine the ensemble shear uncertainty by bootstrap resampling the same size of subsamples of galaxies within the ensemble for $10^4$ times, and adopt the standard deviation of these re-sampled ensemble shear as the errorbar on the shear. When doing so, we confirm that the statistical uncertainty scales like $n^{-\frac{1}{2}}$, where $n$ is the number of 90-degree rotated pairs in the subsample. To reduce the statistical uncertainty of multiplicative bias $m$ below $10^{-4}$, 
we need at least $10^2$ galaxy pairs randomly drawn from the ensemble. The results we show in Section~\ref{sec:great3results} are derived from galaxy ensembles with at least 250 galaxy pairs, so the statistical uncertainty on the shear bias is well below our requirements.

In Fig.~\ref{fig:great3_properties}, we show one- and two-dimentional distributions of galaxy properties for the COSMOS parametric dataset, so as to better understand the population and associated simulation results. The median trend lines in Fig.~\ref{fig:great3_properties} show that with increasing redshift, the galaxies become smaller in apparent size and fainter in magnitude, as expected for a flux-limited population. There is no strong trend in  S\'ersic index as a function of redshift.

\subsubsection{PSF Profile}
\label{sec:PSF_profile}
To simulate errors in the higher moments of the PSF model, we first need to define a base PSF (which  we will refer to as the true PSF). In this work, we generate simulations with \responsemnras{three} parametric base profiles: Gaussian, Kolmogorov, \responsemnras{and Moffat PSF with $\beta = 3.5$.} \footnote{We fit the HSC PSFs with Moffat profiles, and find out the $\beta$ parameter is centered around 3.5. This value is also adopted in \cite{2018MNRAS.481.4445L} on HSC-like simulation.}   Notice that these are the base PSFs rather than the base effective PSFs, which involve the pixel-response function.  \responsemnras{We have another non-parametric PSF: the stacked HSC PSF, which helps us validate the results from the parametric simulations. }

In each simulation, there are two types of PSFs, the model PSF and the true PSF, which are defined using the base PSF with and without any additional kurtosis error.  
To generate the final image, the galaxy profile is convolved with the effective true PSF. 
Both the base true PSF and the base model PSF are round, 
and have the same weighted second moments, given by Eq.~\eqref{eq:secondmoment},  
The shape/shear of the convolved image is estimated separately using the effective true and model PSF.
The shear bias - the difference between these two - is solely caused by the higher moments of the PSF model.

We define model PSFs that differ only in higher moments (not second moments) by perturbing the PSF model differently depending on the original base PSF. For the Gaussian base PSF, we define the model PSF using a S\'ersic 
profile with index $n$ close to (but not precisely) 0.5, since the S\'ersic profile reduces to a Gaussian when $n=0.5$.  For the Kolmogorov base PSF, we construct the model PSF by summing (with equal weights) two Kolmogorov functions with slightly different sizes, parameterized by the ratio of the size of second Kolmogorov to the first. 
\responsemnras{For the Moffat PSF with $\beta = 3.5$, we define the model PSF by varying $\beta$. We are not expecting the results from different types of PSF to be the same, since the higher-moments other than kurtosis are all perturbed differently.}

All three these modifications have two free parameters: one of them (S\'ersic index, Kolmogorov size ratio, or $beta$) is adjusted to explore different modifications to the higher moments of the PSF, while the other is a resizing parameter (S\'ersic and Moffat $R_h$, and the size of the first Kolmogorov) that can be adjusted to achieve our goal of matching the second moments of the model PSF to those of the true PSF.  The actual process of adjusting the resizing parameter to match the observed second moments of model and true PSF is as follows: 
\begin{enumerate}[Step 1:]
    \item  Create the base PSF profile and the initial guess for the model PSF profile, without adjusting the size of the model PSF profile. 
    \item  Convolve both PSFs with the pixel response function and render each into an image with the adopted pixel size.
    \item  
    Measure the observed second moment size $\hat{\sigma}$ of the effective true PSF $\hat{\sigma}_{\text{true}}$ and of the effective model PSF $\hat{\sigma}_{\text{model}}$. Notice that the $\hat{\sigma}_{\text{true}}$ will be slightly different from the assigned $\sigma$ for the base PSF, because of the convolution with the pixel response. 
    \item  Dilate the model PSF base profile by $f'(x') = f'(\hat{\sigma}_{\text{true}} x /\hat{\sigma}_{\text{model}})$,  
    using the expand transformation in \textsc{GalSim}, and replace the old model PSF profile with it. 
\end{enumerate}

Steps 2--4 are repeated until $\hat{\sigma}_{\text{model}} - \hat{\sigma}_{\text{true}} < 10^{-6}$~arcsec. 
Note that with a round base PSF and a model PSF that only differ in the radial moments, the shear bias can only be multiplicative. Therefore, in the rest of the paper, we focus on analyzing the multiplicative bias caused by such modeling error, though future analysis should also consider non-round PSFs and change non-radial moments of the PSF to investigate the additive bias. \response{In the case that the pixel size is comparable to the scale of the PSF, the radial kurtosis values are slightly different when we re-scale the PSF size in Step 4, changing by a few percent. As a result, we cannot strictly control the kurtosis of our model PSF, and we re-measure the actual kurtosis of the model PSF $\hat{\rho}^{\text{(4)}}$ after the transformation to calculate $B[\rakurto]$ by Eq.~\ref{eq:kurtosis_error}.}

\responsemnras{Additionally, we test the shear biases using the stacked HSC PSF directly (these data will be described in Section~\ref{sec:hscdata}). To do so, we interpolate star images and PSFEx models, and stack them with a common centroid, as the true and model PSF respectively. We transform all PSFs so that their shape is round, and they have the same second moment $\sigma$ as the true PSF.  We bin the HSC stars by their kurtosis biases $B[\rakurto]$, producing a true and model PSF for each bin, and measure the shape biases of Gaussian galaxies with different sizes. }

\responsemnras{Although the image simulations in this paper do not include noise, an adaptive weight function that matches the size of the PSF is still applied to the PSF when measuring galaxy shape. This is crucial because (a) it matches how weak lensing shear inference is done in real data, (b) the weight function is the reason why PSF higher moment errors can cause weak lensing shear biases \citep{2011MNRAS.412.1552M}. The choice of the weight function can affect the connection between PSF higher moment errors and shear biases. Therefore, we use the adaptive Gaussian weight function, which adjust its size and shape depending on the PSF, as it is similar to what is effectively used in many moment-based and model-fitting \citep[e.g.,][]{Zuntz:2013zi} shear measurements.   However, we do not explore this nuance in detail as it is beyond the scope of this paper.}

\subsubsection{Simulation Roadmap}
\label{sec:simulation_roadmap}

In this section, we describe the flow of the simulations in this paper. We start with simulations with only one galaxy, in order to isolate the primary factors that determine the shear bias caused by higher moments error of PSF model. Then, we simulate galaxy ensembles with realistic distributions of size, shape and S\'ersic indices to understand the impact on real galaxy surveys. 

The experiments we conduct with single galaxies \responsemnras{and parametric PSFs} are defined in Table~\ref{tab:specification}, and the results are shown in Section~\ref{sec:singleshape}. We gradually increase the complexity of both galaxies and PSFs. First, we simulate a Gaussian galaxy and a Gaussian true PSF, while modifying the galaxy shape to see if the shape bias is multiplicative or additive. We also change the kurtosis biases of the PSF, with several sizes of the galaxies, to check dependency of the galaxy shape biases on the kurtosis biases, described in row~1 of Table~\ref{tab:specification}. We then explore the galaxy size-dependence in greater depth, as described in row~2 of Table~\ref{tab:specification}. Next, we increase the model fidelity for both the galaxies and the PSFs, and conduct the same tests of galaxy size-dependence. For the PSFs, we change the model to Kolmogorov and \responsemnras{Moffat}, in row~3 and \responsemnras{row~4} of Table~\ref{tab:specification}. For the galaxies, we change the model to S\'ersic profiles, and experiment with several values of S\'ersic indices, in row~5 of Table~\ref{tab:specification}. Finally, we conduct experiments with S\'ersic galaxies and Kolmogorov PSFs, in row~6 of Table~\ref{tab:specification}. For rows 2--6, we also use the 90-degree rotated pair method described in Section~\ref{sec:galaxyprofile}, to check whether we can translate the conclusions from shape biases to shear biases. These experiments help us understand the fundamental factors that determine how PSF kurtosis bias translates into galaxy shape and shear bias, which is essential for understanding experiments with higher model fidelity.

Next, we simulate the galaxy ensemble with a realistic distribution of sizes, shapes and S\'ersic indices, obtained using the COSMOS catalog described in Section~\ref{sec:galaxyprofile}, with results shown in Section~\ref{sec:great3results}. The PSF model we use in this experiment is a Kolmogorov profile, with a fixed FWHM of $0.7 \arcsec$. 
We first conduct basic experiments, including changing the kurtosis bias of the PSFs, and the shear of the galaxies, to test the validity of conclusions from the previous, simpler experiments. We also change the size of the galaxy ensembles to understand the errorbars of the shear biases. We investigate the parameter-dependence of the shear bias by creating sub-ensembles binned by particular parameters. We bin the galaxies by their half light radii $R_h$, grouping with/without S\'ersic index. We also explore the redshift-dependence of the induced shear bias by binning the galaxies in redshift bins. We further discuss the consequence of these results for cosmological weak lensing shear measurements in Section~\ref{sec:redshift_dependent_bias}.

\subsection{HSC Data}
\label{sec:hscdata}

In this work, we inspect real data from the Hyper Suprime-Cam \citep[HSC;][]{2018PASJ...70S...4A} to understand how the current level of PSF modeling is doing in recovering higher moments, in specific, the radial kurtosis $\rakurto$. The dataset we are utilizing is the HSC star catalog of the first HSC public data release \citep[PDR1;][]{2018PASJ...70S...8A}.  
The HSC pipeline \citep{2018PASJ...70S...5B} uses a modified version of the \textsc{PSFEx} \citep{2011ASPC..442..435B}, part of the LSST Data Management \citep[DM;][]{2017ASPC..512..279J} codebase, for PSF modeling. 
We use the coadded image of the selected bright stars, 
for which selection criteria will be described in Section~\ref{sec:star_selection}, as the true effective PSF.  These are compared with the coadded PSF models at the same locations as the stars. The details of the PSF modeling and their coaddition in HSC PDR1 can be found in Section~3.3 of \cite{2018PASJ...70S...5B}.  

Below we describe the two key analysis steps applied to HSC data: star selection and kurtosis measurement. 

\subsubsection{Star Selection}
\label{sec:star_selection}

The first step of utilizing the HSC star catalog is to select objects that are suitable for radial kurtosis measurement. First, we apply the first~11 ``basic flag cuts" in table~3 of \citep{2018PASJ...70S..25M},  
and change the \code{iclassification extendedness} to 0 to include only non-extended objects. 
These flags cuts ensure that the coadded images of the objects in our catalog do not include artifacts such as exposure edges, bad pixels, saturation or cosmic rays. The \code{iclassification extendedness} cut is meant to omit extended objects. While our sample may still include small galaxies that are classified as non-extended,  \cite{2018PASJ...70S...5B} showed that the classification works well for objects brighter than $i\sim 24$, which describes the star sample we are using. 

We also determine the minimum signal-to-noise ratio (SNR) cut on the stars so as to ensure the measurement of the radial weighted kurtosis has a reasonable statistical precision for our purposes. To set a SNR threshold, we simulate stars with $10^4$ noisy realizations of the same profile with a certain SNR with \textsc{GalSim}. We then measure the radial kurtosis of all realizations to estimate the relationship between statistical uncertainty on radial kurtosis and SNR. The results of this exercise suggest that to achieve a statistical uncertainly $\delta B[\rakurto] \lesssim 0.1$\% in the radial weighted kurtosis, the SNR should exceed 800.  Therefore, we set the minimum SNR to 1000. 
The 0.1\% threshold still leaves us with a reasonably-sized sample while ensuring sufficiently high-precision measurements.

The final step in generating a star 
catalog is removing objects whose light profiles are contaminated by light from other objects. Given the sensitivity of the radial kurtosis to the outer part of the light profile, this step is particularly important. We do this using two methods: removing double stars and removing blended stars. To remove double stars, we detect them by scanning through the entire catalog of objects flagged as unique detections with the \code{idetect\_is\_primary} flag using a k-d tree structure. 
With a k-d tree, we can detect any two objects in the entire catalog that are located within some chosen tolerance (here we choose 2\arcsec), and call them ``objects with near neighbor(s)". 
We then remove any stars with such near neighbor(s) detected from our star catalog. 

\begin{figure}
    \centering
    \includegraphics[width = 0.99\columnwidth]{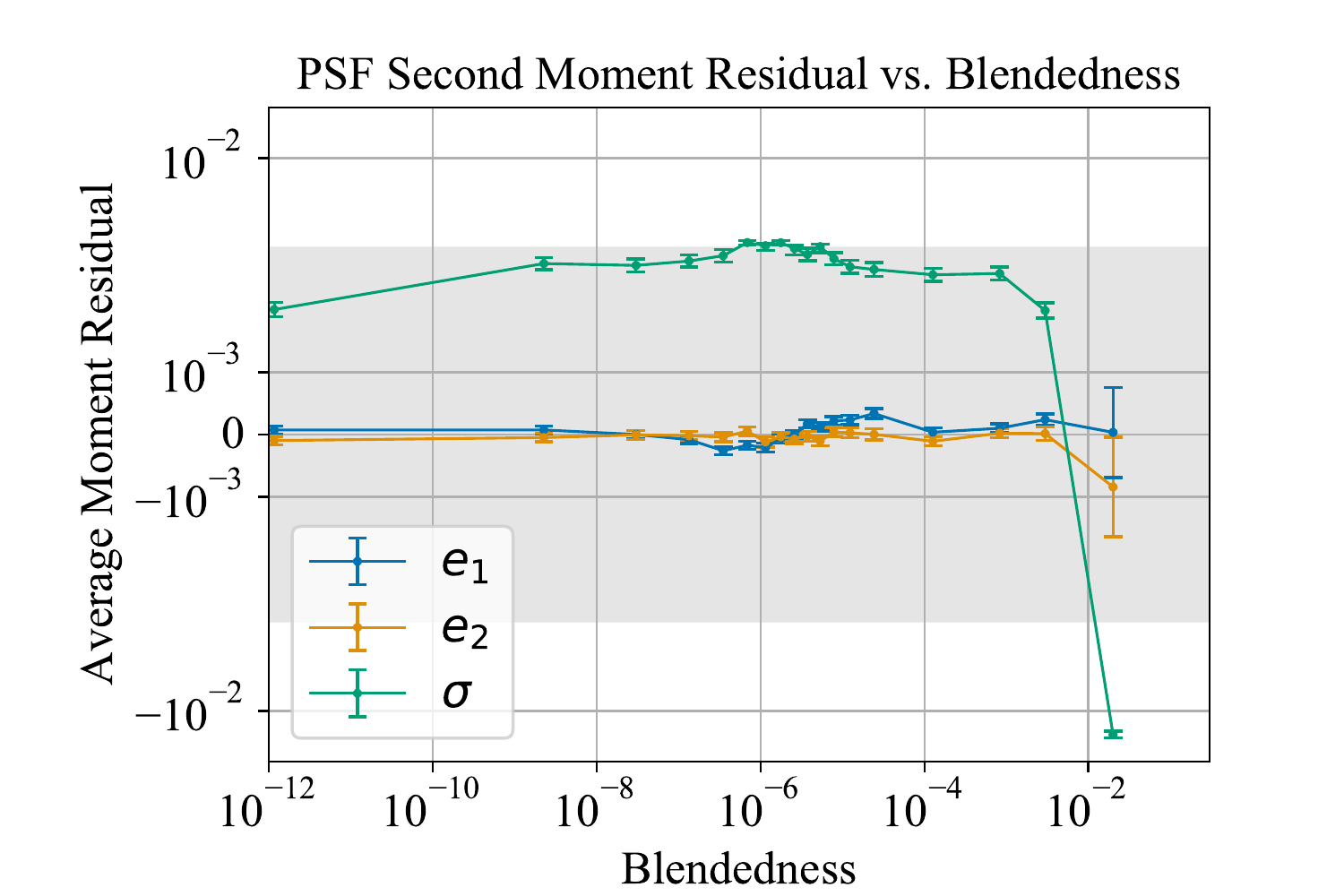}
    \caption{Average second moments residual (comparing the moments of the PSF model to those measured in the postage stamp image) for bright stars in the HSC survey, binned by the blendedness of the star. When the blendedness is larger than 0.001, we see significant residuals in the second moments. \response{The y-axis is symmetric-log scaled with a linear threshold $=0.003$. The linear region is shaded. }
    }
    \label{fig:hsc_blendedness_cut}
\end{figure}

To remove the blended objects, we utilize the parameter \code{iblendedness\_abs\_flux}, which describes how much flux of the object is potentially from other  
objects, and set an upper limit based on tests to determine when blending may be affecting the image enough to noticably impact the second moments. We use the residuals (difference between PSF model and moments measured from the image) of the second moment $\sigma$, $e_1$ and $e_2$ for this purpose. In Fig.~\ref{fig:hsc_blendedness_cut}, we show the PSF model residuals of stars in our catalog before applying a cut to remove blended objects, binned by the blendedness of the stars. We can see that when the blendedness exceeds $10^{-3}$, the second moments of the bright stars measured from the images differ noticeably from the moments of the PSF model. Therefore, we exclude stars with \code{iblendedness\_abs\_flux} $>10^{-3}$. \response{We also notice a positive bias on second moment $\sigma$ across all blendedness bins. This is likely connected to the brighter i-band magnitude of the stars we have selected:  \citet{2018PASJ...70S...5B} shows that brighter PSF stars tend to have positive $\delta \sigma / \sigma$.}

\begin{table}
\begin{tabular}{ccc}
\hline
Steps & Criteria                & Number of Objects\\\hline
1     & Basic flags cuts        & $2.7\times 10^6$            \\
2     & Signal-to-noise $>10^3$ & $8.0 \times 10^4$           \\
3     & Blendedness $<10^{-3}$  & $7.0 \times 10^4$           \\
4     & Nearby object detection & $6.8 \times 10^4$       \\\hline
\end{tabular}
\caption{The number of stars remaining after each operation in our star selection on the HSC \code{GAMA\_15H} field. The details and reasoning for the cuts are explained in Section~\ref{sec:star_selection}.}
\label{tab:star_cuts}
\end{table}

With these two methods, we remove stars that are too close to other (likely compact) objects, or are contaminated by a potentially extended background light profile. Out of the six fields in the HSC dataset, we chose to analyse a field with better-than-typical seeing, 
since the 
better-seeing fields generally have worse PSF modeling quality \citep{2018PASJ...70S..25M}, and are better samples for testing the PSF model. 
We choose \code{GAMA\_15H}, since it has the best seeing among the fields that pass the nominal cuts on PSF modeling quality in \citet{2018PASJ...70S..25M}. 
The number of stars remaining after each cut is applied to the catalog is shown in Table~\ref{tab:star_cuts}.  
At the end of the selection process, we have $\sim$ 68,000 star samples  
for investigating the PSF modeling quality.

\response{We use the same flag cuts for stars as in \citet{2018PASJ...70S..25M}; however, the SNR cut that we add to the flag cuts results in selection of a star population that is brighter than that in \citet{2018PASJ...70S..25M}. Our selected stars have i-band magnitudes ranging from $18$--$20.5$. As a result, the second moment distribution is slightly larger than that in \citet{2018PASJ...70S..25M}, but the dataset can still serve the purpose of this paper.}

\subsubsection{Analysis of HSC stars}

After we select the stars to create the catalog, we retrieve postage stamp images of the stars (which we consider as representing the true PSFs), and the model PSFs reconstructed at the positions of the stars. We are using the coadded images rather than the original individual exposures, to which CCD-level processing was applied. The PSF models are also appropriately weighted coadditions of the individual exposure PSF models.
We measure the second moments and radial kurtosis of the stars and model PSFs to obtain a catalog of stars with moments of their true and model PSF, as a function of  their position on the sky. 

\responsecwr{The magnitude distribution of the selected stars is within but at the brighter side of the i-band magnitude distribution for PSF stars in \cite{2018PASJ...70S..25M}.}  We also inspect the second moment modeling quality of our selected stars. We use 
\begin{equation}
    f_{\delta \sigma} = \frac{\sigma_{\text{model}} - \sigma_{\text{true}} }{\sigma_{\text{true}}}
\end{equation}
to measure the size model quality. We bin our selected star by their i-band magnitude into 10 bins, and find that the average $f_{\delta \sigma}$ values for each bin do not exceed the requirement on $f_{\delta \sigma}$ for HSC, $0.004$, reproducing the results of Fig.~6 in \cite{2018PASJ...70S..25M}.

This catalog enables us to determine the mean value and standard deviation of the true and residual PSF kurtosis, where the residual PSF kurtosis is defined by Eq.~\eqref{eq:kurtosis_error}. 

\responsemnras{For a given sample of galaxies used to measure the weak lensing shear, the shear field depends on the position on the sky $\mathbf{x}$; so does the kurtosis bias, and therefore the associatd shear multiplicative bias.  The observed shear $\hat{g}(\mathbf{x}) = [1+m(\mathbf{x})] g(\mathbf{x})$, where the $g(\mathbf{x})$ is the true shear.  When $|m| \ll 1$, the observed shear correlation function is
\begin{align}
    \nonumber\langle \hat{g}(\mathbf{x}) \hat{g}(\mathbf{x+\theta})\rangle = & (1 + 2\langle m\rangle) \langle  g(\mathbf{x}) g(\mathbf{x+\theta})\rangle.
\end{align}
The mean multiplicative bias of a galaxy ensemble $\langle m\rangle$ can be calculated by the average multiplicative bias of the galaxies in it, $\langle m \rangle = \langle m(B[\rakurto]) \rangle$, when there is no other source of systematics. \response{Since the shape and shear biases are proportional to the kurtosis bias, as we later find out, we can estimate the shear bias by taking the first term of Taylor expansion
\begin{equation}
\label{eq:m_estimate}
    m(B[\rakurto]) \approx \frac{\partial m}{ \partial B[\rakurto]} B[\rakurto].
\end{equation}
}The first factor in the equation above is determined primarily by the galaxy population and the second factor by the PSF.  So, they are independent random variables and the averages can be separately calculated.
We have tested this linear approximation and found it to be accurate at the level of $\sim 0.1\%$ of the measured shear bias, within the $B[\rho^{(4)}]$ range of HSC data. }

\responsemnras{For a galaxy ensemble, based on the assumption of Eq.~\ref{eq:m_estimate}, $\langle m \rangle = \langle \partial m/ \partial B[\rho^{(4)}] \rangle \langle B[\rho^{(4)}] \rangle $.  Functionally, this means that we can simply calculate the average of $ B[\rho^{(4)}] $ over the PSF model across the survey, and average $\partial m/ \partial B[\rho^{(4)}]$ over the galaxy population.  Those two separate results can be combined to estimate an average weak lensing shear bias for galaxy populations that resemble those that will be used for measurements of weak lensing surveys such as LSST.}

\section{Results}
\label{sec:results}

In this section, we show the results of carrying out the measurements described in Sec.~\ref{sec:methods}. First, we show the results of the image simulation -- simulations with simpler galaxy populations in Section~\ref{sec:singleshape} and using the full COSMOS catalog in Section~\ref{sec:great3results}. Next, we show the results of analyzing the moments of the HSC PDR1 star sample in Section~\ref{sec:realPSF}. Finally, we estimate the redshift-dependent weak lensing shear bias caused by errors in the higher moments of the PSF  in Section~\ref{sec:redshift_dependent_bias}, combining the simulation and HSC results.

\subsection{Single Galaxy Experiments}
\label{sec:singleshape}

Here we show the results of controlled numerical experiments that test the impact of errors in the higher moments of the PSF model on the shape measurement of a single galaxy.

First, we check the behavior of the shape measurement bias $\delta e = \hat{e} - e$ 
caused by the PSF kurtosis bias. We simulate a Gaussian galaxy convolved with a round Gaussian PSF, which has kurtosis $\rho^{(4)}=2$, as shown in Table~\ref{tab:kurtosis}. 
The model PSF is generated using a  S\'ersic profile with index slightly different from $n=0.5$ (Gaussian case) as mentioned previously in Section~\ref{sec:PSF_profile}. We simulate several galaxies with different ellipticities and find out that the shape error $\delta e$ is proportional to the galaxy shape $e$, 
which means that the shape error caused by kurtosis is a multiplicative bias.  \response{We also carried out a test with $e = 0$ to verify that the additive bias on the shape is zero under PSF kurtosis bias, for the round PSF configurations used in this experiment. } Since the shape bias is multiplicative, in future experiments regarding shape error, we only simulate one value of \response{$e = (0.28, 0.0)$, i.e. the intrinsic galaxy shape dispersion,} and use $\delta{e}/e$ as the multiplicative bias.  

\begin{figure}
    \centering
    \includegraphics[width=0.98\columnwidth]{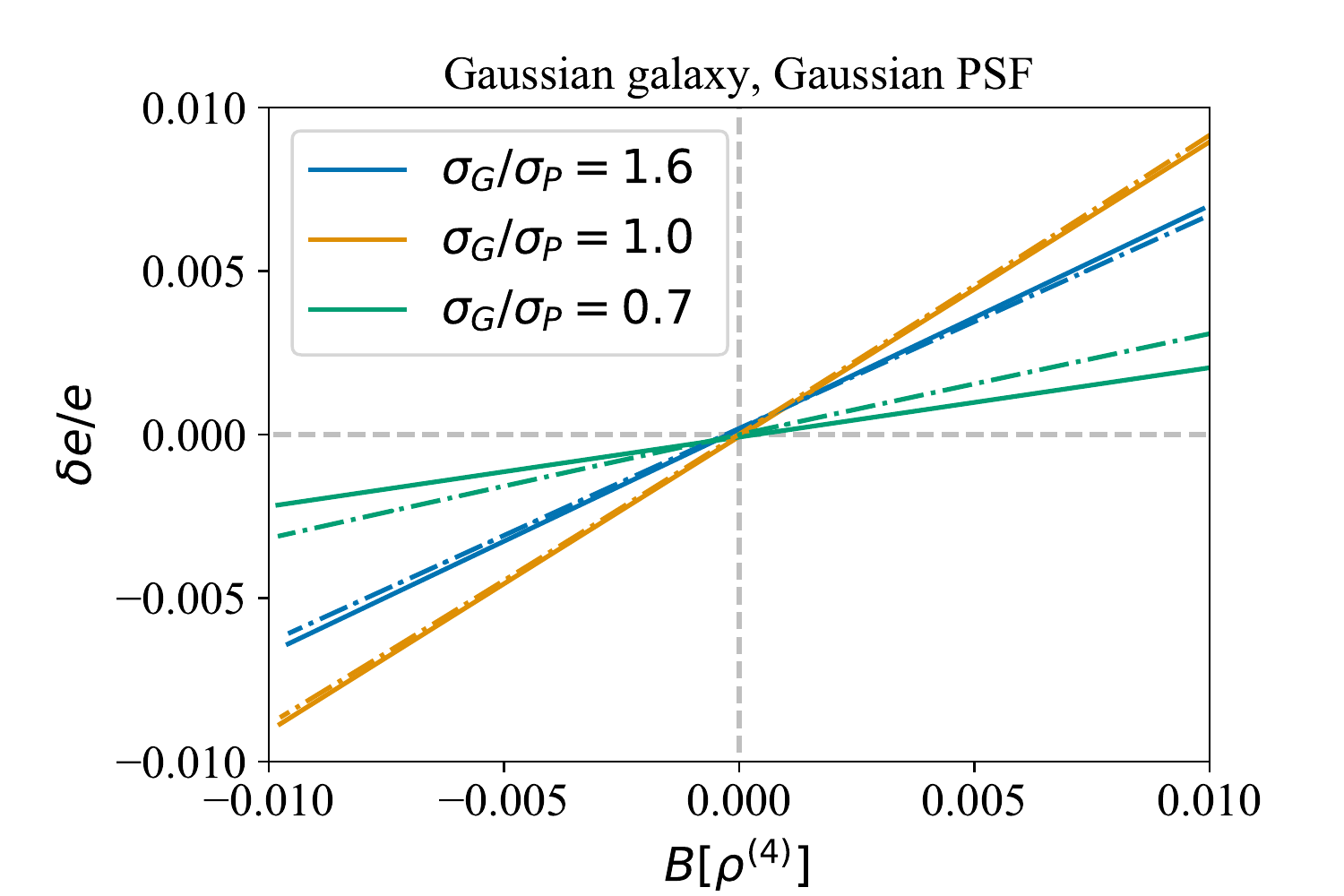}
    \caption{Here we show the galaxy shape bias as a function of PSF kurtosis bias for a Gaussian galaxy and Gaussian PSF, with re-Gaussianization (\response{dot-dashed}) and metacalibration applied to re-Gaussianization (\response{solid}) for three different size ratios of the galaxy and PSF. Dashed lines indicate the zero value for both plotted quantities. As shown, the galaxy shape bias depends linearly on the PSF kurtosis bias, and depends in a more complex way on the galaxy versus PSF size ratio. The kurtosis bias is defined in Eq.~\eqref{eq:kurtosis_error}.  
    }
    \label{fig:GGlinear}
\end{figure}

In the second experiment, we simulate Gaussian galaxies with a single value of $e$ and $\sigma$, and convolve them with a Gaussian PSF, of which the values are shown in row~1 of Table~\ref{tab:specification}.  We then measure the shape of the galaxy using the PSF model that has a perturbing S\'ersic index around 0.5, and with the same second moments as the true PSF, explained in  Section~\ref{sec:PSF_profile}.  
Compared to the last experiment, we are changing the amount of kurtosis error in our model PSF, to determine the relationship between galaxy shape bias and PSF kurtosis error.   
In Fig.~\ref{fig:GGlinear}, we show that the multiplicative galaxy shape bias for a single galaxy is linearly proportional to the kurtosis bias, for both re-Gaussianization and Metacalibration. \response{In later image simulations, we focus quantifying $\partial m / \partial B[\rakurto]$, so that we can predict the shear bias for the HSC dataset by combining the simulation with the measurement of $B[\rakurto]$, as shown in Eq.~\eqref{eq:m_estimate}.} This constant of proportionality depends on the ratio of the size of the galaxy to the size of the PSF (and, notably, is not monotonic in that ratio). Our next goal is to explore the potentially complex dependency on the ratio of galaxy-to-PSF sizes. 

\begin{figure}
\centering
\includegraphics[width=0.98\columnwidth]{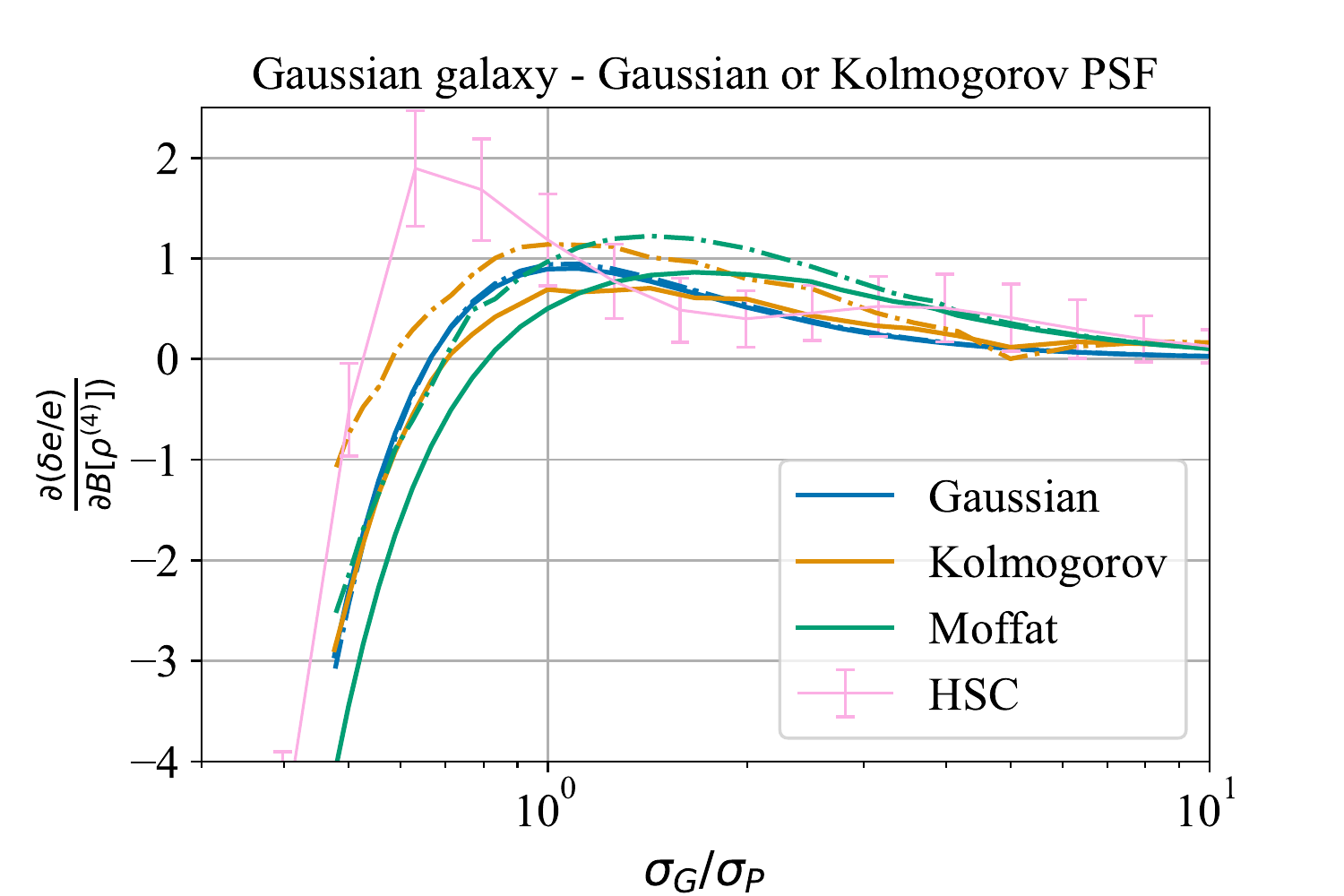}
\caption{This plot illustrates the relationship between the slope of the lines from Fig.~\ref{fig:GGlinear} (ratio of galaxy shape bias to kurtosis bias) and the size ratio between the galaxy and PSF.  The colors of the lines indicate the functional form for the true PSF (indicated in the legend). Dot-dashed lines show the results when using metacalibration, and the solid lines show the results for re-Gaussianization. \responsemnras{The stacked HSC PSF are only measured in re-Gaussianization.}  
As shown, the trends in the dependence on the galaxy versus PSF size ratio are quite similar for \responsemnras{all four} PSF models and shape measurement methods. 
}
\label{fig:GGsize_ratio}
\end{figure}

\begin{figure*}
    \centering
    \includegraphics[width=0.9\columnwidth]{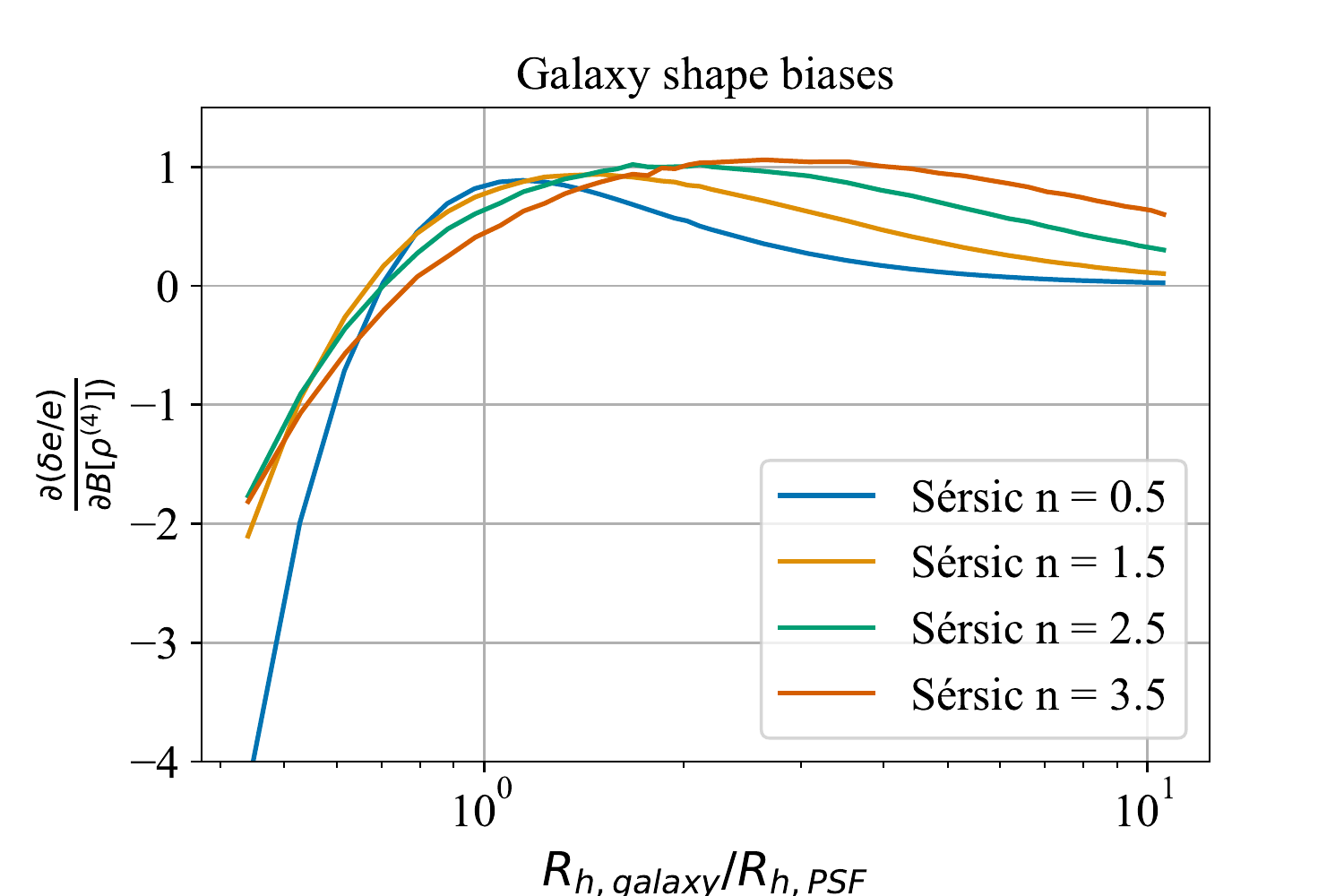}
    \includegraphics[width=0.9\columnwidth]{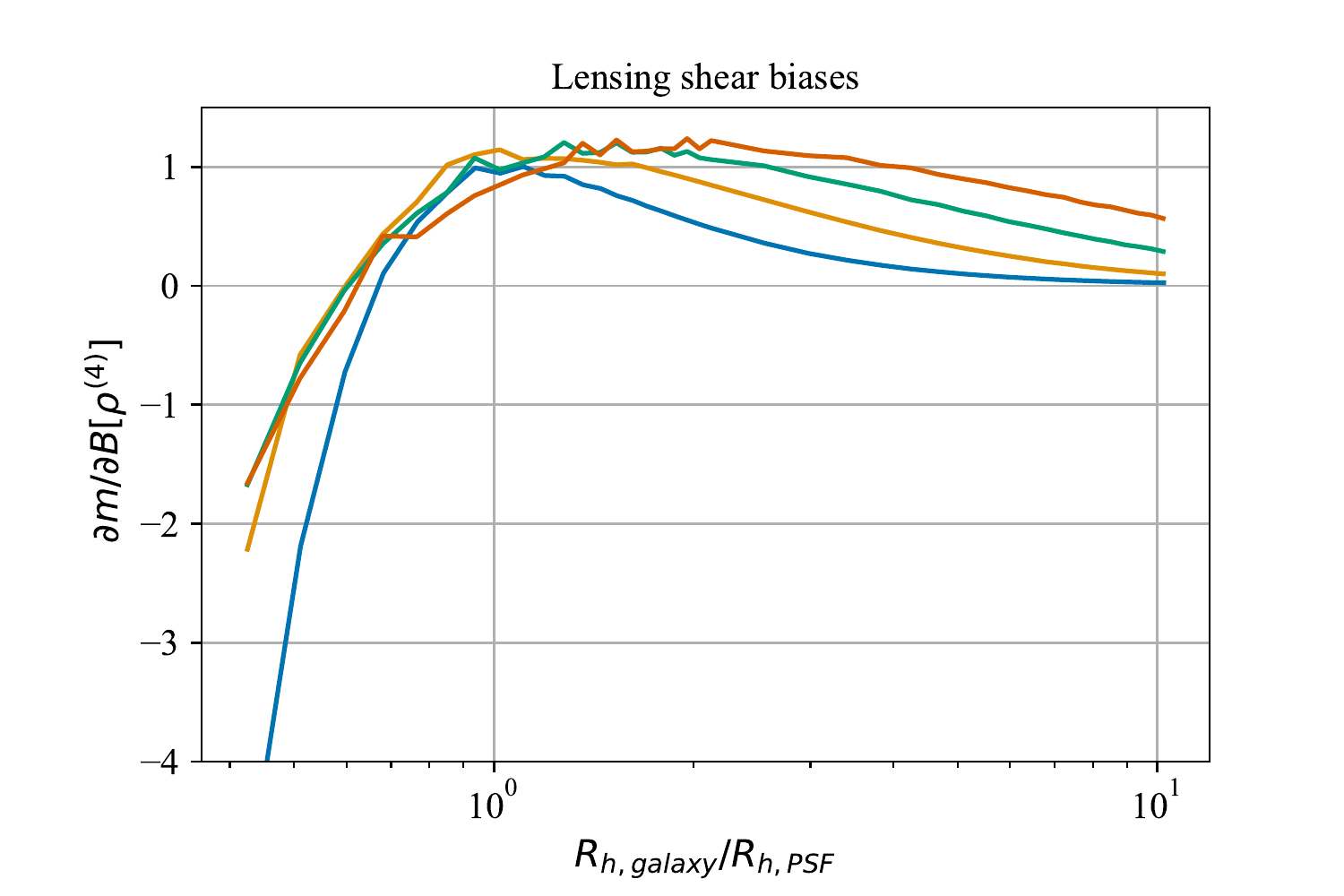}
    \caption{\label{fig:Kolmogorov_combine}
    \textbf{Left:} The relationship between the ratio of galaxy shape bias to PSF kurtosis bias and the galaxy-to-PSF size ratio, simulated with a S\'ersic galaxy profile and  Gaussian PSF. 
    \textbf{Right:} Same as the left panel, but for the ensemble shear bias (rather than galaxy shape bias), measured using 90-degree rotated pairs and a Gaussian PSF.     
    The measurements in both panels are made using re-Gaussianization. As shown, both the galaxy shape and weak lensing shear bias are only mildly dependent on the S\'ersic index, with the minor differences between the curves for different S\'ersic indices being subdominant to the dependence on galaxy-to-PSF size ratio. 
    }
    
\end{figure*}

\begin{figure*}
    \centering
    \includegraphics[width=0.45\textwidth]{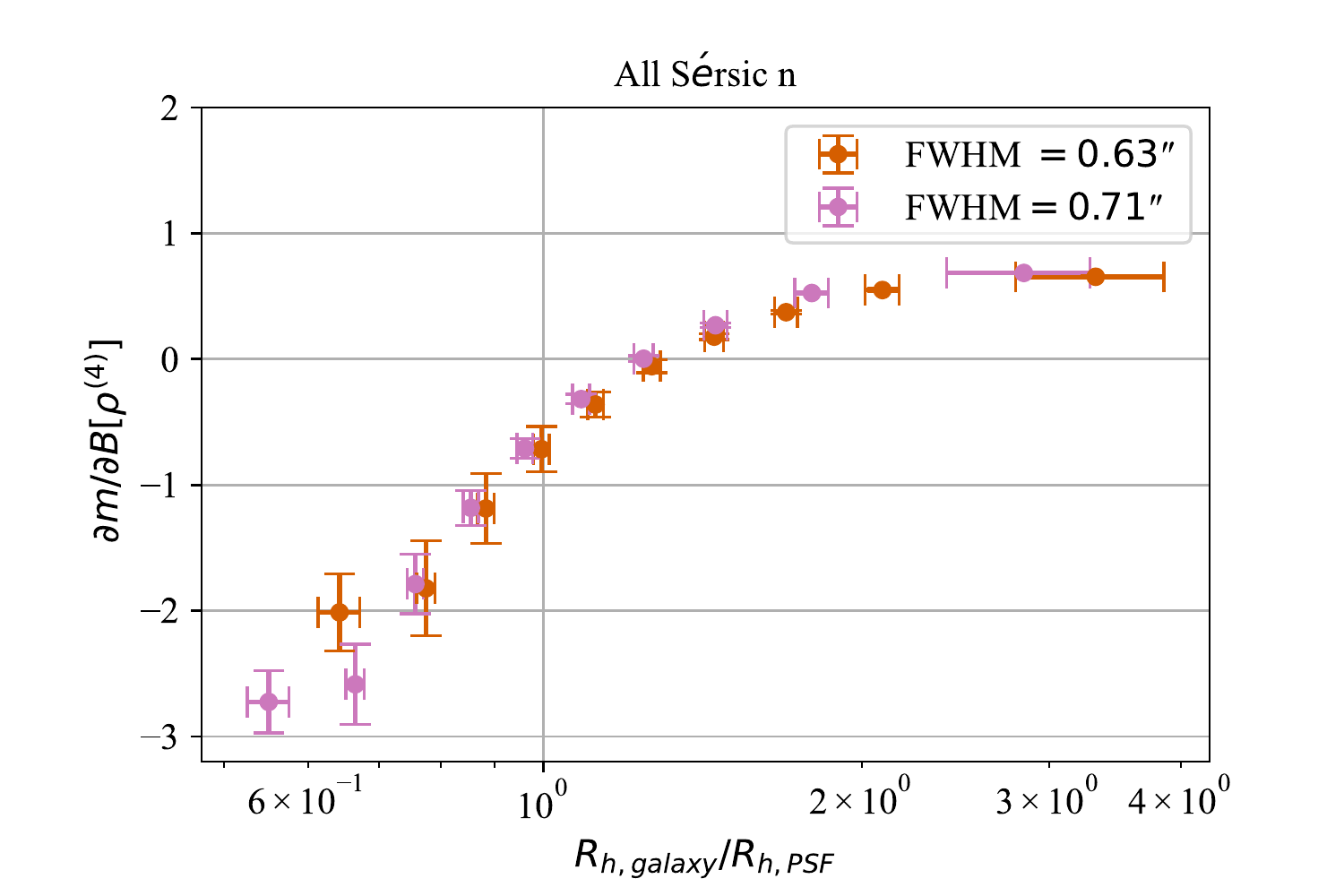}
    \includegraphics[width=0.45\textwidth]{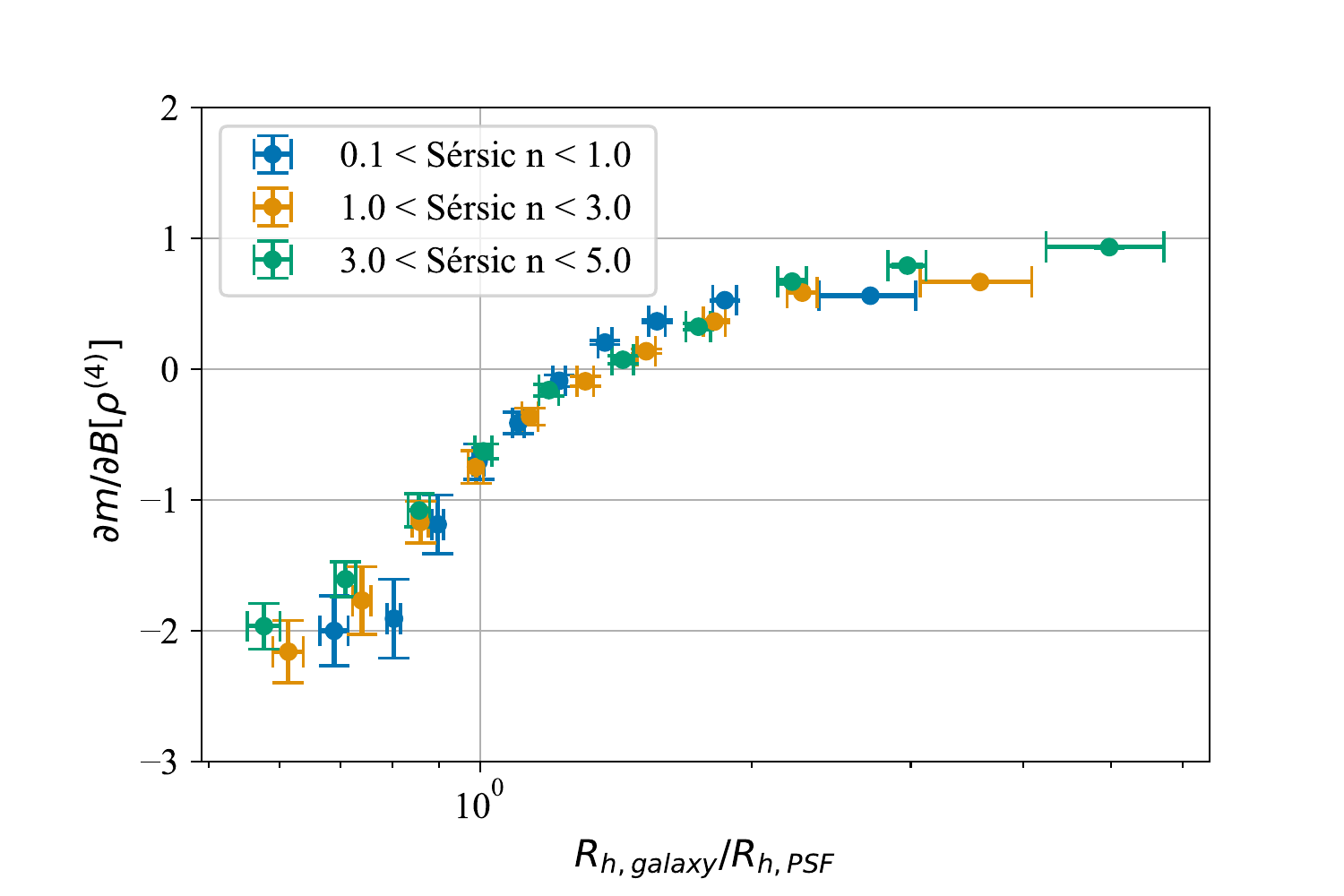}
    \caption{\textbf{Left:} Multiplicative bias per PSF kurtosis bias for subsamples of galaxies from the COSMOS parametric catalog, \response{binned by the size ratio $R_{h,\text{galaxy}}/R_{h,\text{PSF}}$, including all S\'ersic indices, for two runs with PSF FWHM $=0.63$ and $0.71$.} \textbf{Right:} \response{The same quantities as on the left}, shown separately for three ranges of S\'ersic index. The variation between the curves with different S\'ersic indices is significantly less than the variation with galaxy size, just as for single galaxy simulations. The horizontal errorbars show the standard deviation 
    within each bin, while the vertical errorbars show the uncertainty on the mean value, generated by bootstrap resampling from the fixed subsample of the COSMOS catalog $10^4$ times.
    \label{fig:great3_bin_property}
    }
\end{figure*}

The results of exploration of the relationship between the galaxy shape bias and the kurtosis bias, expressed in terms of the derivative $(\partial \hat{e}/e)/(\partial B[\rakurto])$, are shown in Fig.~\ref{fig:GGsize_ratio}.  
In this test, we simulate Gaussian galaxies \responsecwr{with three parametric PSFs: Gaussian PSFs (blue), Kolmogorov PSFs (orange), \responsemnras{Moffat PSFs (green)}, with shear estimation via re-Gaussianization (solid) and metacalibration (dot-dashed).} The main parameters of this experiment are specified in rows~2-4 of Table~\ref{tab:specification}. \responsemnras{In addtion, we also measure the shear estimation with a non-parametric PSF: the stacked star images and PSFs from the HSC data.} The galaxies have a shape of $e_1 = 0.28$ and $e_2 = 0$. \responsemnras{We find that the shape biases respond to different PSFs with a similar trend.}  We tested other \responsecwr{moment-based} shape measurement methods, and we found that \responsecwr{the responses from these shape measurements follow the same trend as a function of the size ratio, though potentially differing in magnitude by a factor up to 2 for small galaxies.} \responsecwr{In \cite{2020A&A...636A..78S}, shape measurement methods seem to have very different responses to the HME.}
However, there are a few differences between the two studies: (a) the tests in \cite{2020A&A...636A..78S} are for a space-based telescope, while we study a ground-based telescope; (b) \cite{2020A&A...636A..78S} also explore more complicated PSF model residuals, while we have controlled experiments that only have radial moment residuals; \responsecwr{and (c) the shear measurement methods in \cite{2020A&A...636A..78S} also have more fundamental differences from each other, while the shape measurement methods compared in this study are relatively similar}.

Next, we compare the metacalibration results to those for re-Gaussianization; both are shown in Fig.~\ref{fig:GGsize_ratio}. \responsecwr{Metacalibration applied to re-Gaussianization has a similar results as  re-Gaussianization alone for the Gaussian PSF. However, for the Kolmogorov PSF, we do see differences between the two methods of up to a factor of 2, despite the fact that the qualitative trends with galaxy-to-PSF size ratio are similar. }  

In Fig.~\ref{fig:GGsize_ratio}, we also add complexity to the PSF model in the simulation, specified in row~3 and row~4 of Table~\ref{tab:specification}.  We use a Kolmogorov profile as the true PSF, and the double-Kolmogorov as a perturbed model PSF, again with preserved second moments. \responsemnras{We also use the Moffat PSF with $\beta_0 = 3.5$ as the true PSF, and vary the $\beta$ as the perturbed PSF.} This figure therefore provides a comparison among results with \responsemnras{Gaussian, Kolmogorov, Moffat PSF and stacked HSC PSF -- in all four cases, with a Gaussian galaxy.} We find that compared to a Gaussian PSF, kurtosis residuals in the model for the Kolmogorov PSF and \responsemnras{Moffat} cause slightly more shape bias given the same kurtosis bias, for larger galaxies. While it may seem that the shape bias in the case of a Kolmogorov and \responsemnras{Moffat} PSF converges to a positive constant, this is not the case: we have confirmed that the shape bias converges to zero for sufficiently large galaxies with $\sigma_G/\sigma_P > 20$. \responsemnras{ It is expected that the results from different types of PSF are different by a factor-of-a-few, since a fixed kurtosis bias corresponds to different perturbations of the other higher moments. This is especially true for the real stacked PSF, as kurtosis biases might be correlated with other higher-moments biases in real data. However, the goal of this paper is to provide an initial  order-of-magnitude estimate of the impact of errors in the higher order moments of the PSF.  At that level, all our results show a consistent magnitude and dependence on galaxy and PSF size ratio, which points to kurtosis bias as the most important higher moment for determining the weak lensing multiplicative shear bias. }

Next, we extend our results to greater complexity in the galaxy model by using S\'ersic profiles, which have one more parameter than Gaussian profiles. 
Since the galaxy shape bias is multiplicative and is directly proportional to the PSF kurtosis bias, we present the results in the same form as Fig.~\ref{fig:GGsize_ratio}. The parameters of this experiment are specified in row 4 of Table~\ref{tab:specification}. The left panel of  Fig.~\ref{fig:Kolmogorov_combine}  
shows the results of simulating a series of S\'ersic profile galaxies with S\'ersic indices $n$ ranging from 0.5--3.5, and with galaxy-to-PSF size ratio ranging from 0.5--3. We use the half light radius ($R_h$) to define the size ratio between the galaxy and PSF in this case. 
The S\'ersic galaxy with $n = 0.5$ is simply a Gaussian galaxy, and the result for that case is the same as in Fig.~\ref{fig:GGsize_ratio}. 
As shown, the S\'ersic index plays a relatively minor role in determining the galaxy shape bias for a given level of PSF kurtosis bias.  However, this result would not hold if we had plotted the results as a function of second moment size, since galaxies with the same $\sigma$ and different S\'ersic indices have quite different half light radii. \response{If we use the second moment $\sigma$ as the scale parameter, the different S\'ersic index curves would unify at large size ratios ($\sigma_{\text{galaxy}}/\sigma_\text{PSF} > 2$), but would be highly discrepant for small size ratios. Since most of the galaxies that we are interested in have a small size ratio, we choose to use the half light radius $R_h$ as the scale parameter.  }

Our final step in this section is to switch to measuring the ensemble weak lensing shear bias (rather than galaxy shape bias) due to errors in the PSF higher moments. We simulate a galaxy with $e_1 = 0.28$ 
and its 90-degree rotated pair. \response{Again, we check that the additive bias is zero when $g = 0$. We then apply non-zero shear to check the multiplicative biases. }
In the right panel of Fig.~\ref{fig:Kolmogorov_combine}, we simulate S\'ersic galaxy pairs convolved with a Gaussian PSF, with the shear of the 90-degree rotated pair measured as described in Section~\ref{sec:shapemeasurement}. 
The shape measurement method used in this experiment is re-Gaussianization. 
This plot shows that the ensemble shear bias induced by PSF kurtosis bias is nearly the same as the induced galaxy shape bias, indicating that we can generalize the results from earlier in this section to shear bias. We also test with Kolmogorov PSF, and receiving results \responsecwr{with similar trend to} Gaussian PSF. In the rest of this work, we will focus on tests of ensemble shear recovery with galaxy ensembles, using the ensemble shear multiplicative bias $m$.

\subsection{Experiments with Realistic Galaxies}
\label{sec:great3results}

Here we extend the results from Section~\ref{sec:singleshape} on weak lensing shear bias due to PSF model kurtosis bias for individual galaxies (as a function of their properties) to consider a galaxy population with a realistic distribution of galaxy sizes, shapes, and S\'ersic indices. The galaxy population we use is based on the COSMOS parametric catalog, the galaxy cut is described in Section~\ref{sec:galaxyprofile}. \response{For the following tests we use a Kolmogorov PSF, run twice with $\text{FWHM} = 0.63$\arcsec\ and $\text{FWHM} = 0.71$\arcsec,} and the PSF model is a double-Kolmogorov PSF as described in Section~\ref{sec:PSF_profile}. We impose a cut on the resolution factor $R_2$, according Section~\ref{sec:galaxyprofile}. The ensemble shears are measured by Metacalibration.

Before proceeding based on the assumptions from single galaxy experiments, we confirmed the following conclusions from the previous subsection carry over to ensembles of galaxies with varying sizes and shapes: the weak lensing shear bias generated by PSF kurtosis bias is multiplicative (proportional to the shear) and  proportional to the kurtosis bias. 
We test these conclusion for shear $|g|<0.01$ and for kurtosis bias $|B[\rakurto]| < 0.004$ . This enables us to continue quantifying our results in terms of the multiplicative shear bias per PSF kurtosis bias, or $\partial m/ \partial B[\rakurto]$. In the later simulations, both shear and kurtosis bias are kept constant with $(g_1, g_2) = (0.01,0.0)$ and $B[\rakurto] = 0.0015$. The vertical errorbars of the results in this section is determined by the bootstrap resampling method discussed in Section~\ref{sec:galaxyprofile}, and the horizontal errorbars are the standard deviation of the binned properties.

We want to confirm that the dependence of the ensemble shear bias on the size ratio of galaxy and PSF still holds for the ensemble. On the left panel of  Fig.~\ref{fig:great3_bin_property}, we bin the entire catalog in equal number of galaxies by the \response{half light radius size ratio} of the galaxy \response{over PSF}, and show the relationship between multiplicative shear bias per kurtosis bias and the \response{size ratio for both $\text{FWHM} = 0.63$\arcsec\ and $\text{FWHM} = 0.71$\arcsec.}  We show that the primary determining factor for shear bias induced by PSF kurtosis bias is \response{the size ratio between the galaxies and the PSFs, as previously shown in the simpler experiments in  Sec.~\ref{sec:singleshape}}. 
We further confirm that the results in Section~\ref{sec:singleshape} can be generalized to galaxy ensemble shear by splitting the galaxies in the COSMOS catalog based on their S\'ersic indices, shown in the right panel of Fig.~\ref{fig:great3_bin_property}. \response{For this test, we only plot the results from $\text{FWHM} = 0.63$\arcsec\ run.} The three sets of galaxies have similar shear biases despite the fact that the S\'ersic indices differ significantly for the three groups -- a similar conclusion as from Fig.~\ref{fig:Kolmogorov_combine} (right panel).

\begin{figure}
    \centering
    \includegraphics[width=0.98\columnwidth]{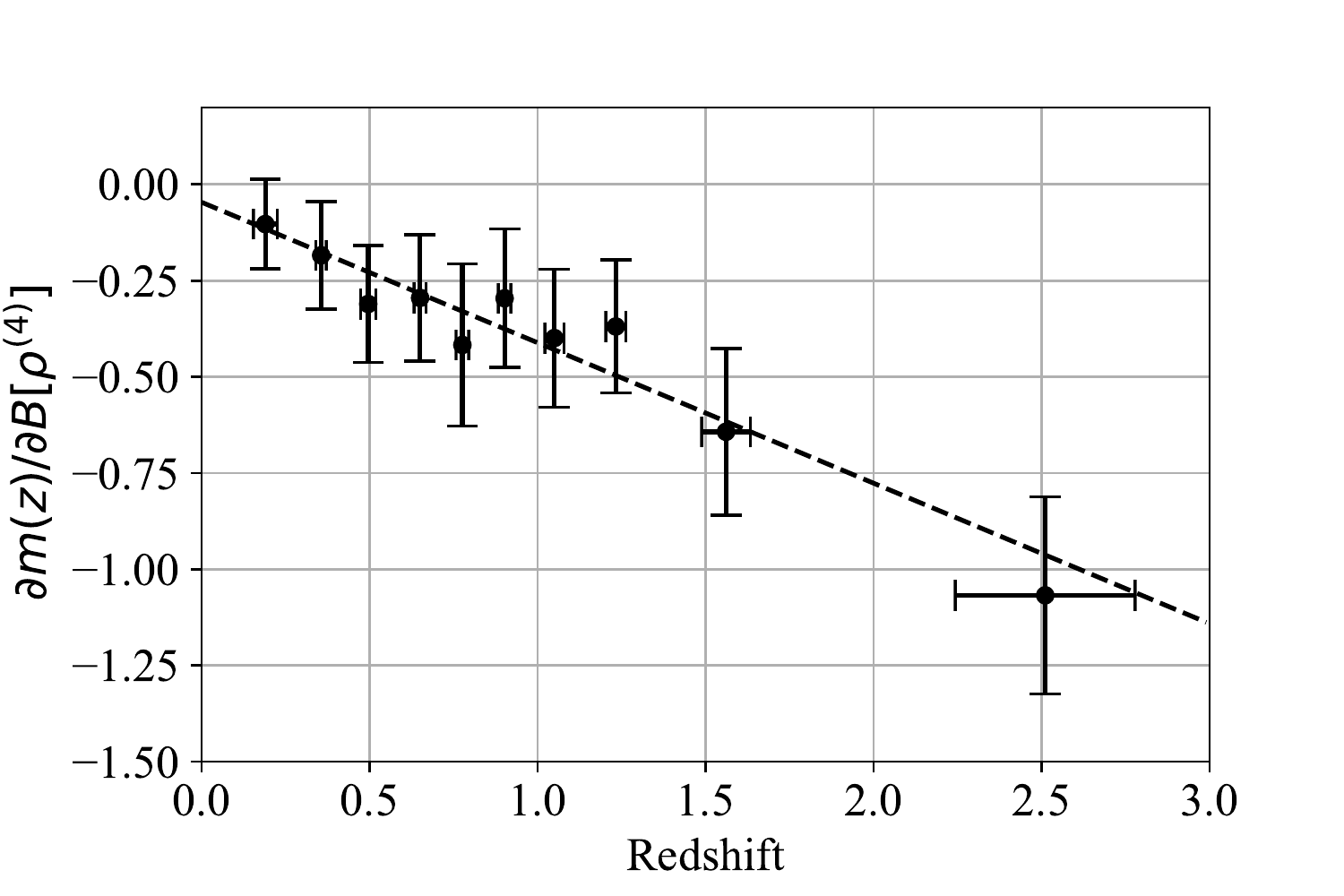}
    \caption{\label{fig:derived_quant}Ratio of weak lensing shear bias and PSF kurtosis bias when binning the COSMOS parametric galaxies by their photometric redshift. The effect can by explained by the fact that galaxies at higher redshift tend to be smaller in apparent size, which results in a more negative shear bias for a given value of PSF kurtosis bias. 
    The horizontal error bar shows the standard deviation within each redshift bin, while the vertical errorbar shows the error on the mean value, generated by bootstrap resampling from the redshift bin $10^4$ times. The dashed-line shows the linear model specified by Eq.~\eqref{eq:m0}.
    }
\end{figure}

Since weak lensing cosmology analyses typically involve tomography, i.e., binning by redshift, we have a strong motivation to investigate what happens to the ratio of shear bias and PSF kurtosis bias when binning the COSMOS parametric galaxies by other properties such as redshift. The results of this experiment are shown in Fig.~\ref{fig:derived_quant}.  
We show that $\partial m/ \partial B[\rho^{(4)}]$ becomes more strongly negative at higher redshift.  This can be explained in terms of the trend in Fig.~\ref{fig:great3_properties}, which showed that galaxy sizes are smaller at higher redshift, and Fig.~\ref{fig:great3_bin_property}, which showed that smaller galaxies have a more negative value of $\partial m/ \partial B[\rho^{(4)}]$.  
As \cite{2013MNRAS.429..661M} noted, the inferred dark energy equation of state is relatively insensitive to a constant multiplicative bias $m_0$. Rather, redshift-dependent multiplicative bias $m(z)$ can more directly mimic changes in the dark energy model \citep{2020MNRAS.496.5017G}.  This means that we need to properly model the redshift-dependent shear bias caused by kurtosis to ensure unbiased cosmological parameter constraints.

\subsection{HSC PSF Modeling}
\label{sec:realPSF}

\begin{figure*}
    \centering
    \includegraphics[width=2.0\columnwidth]{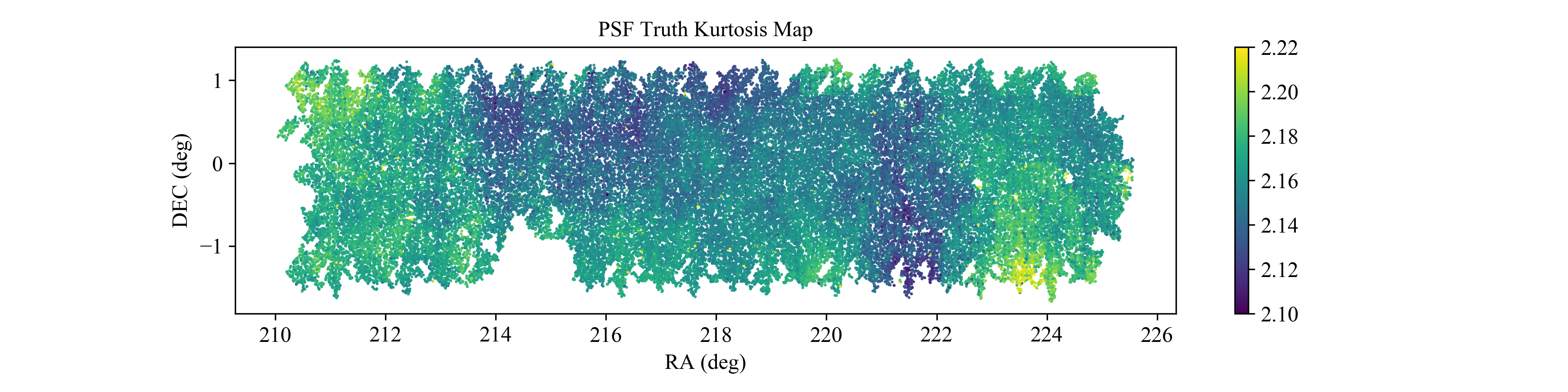}
    \includegraphics[width=2.0\columnwidth]{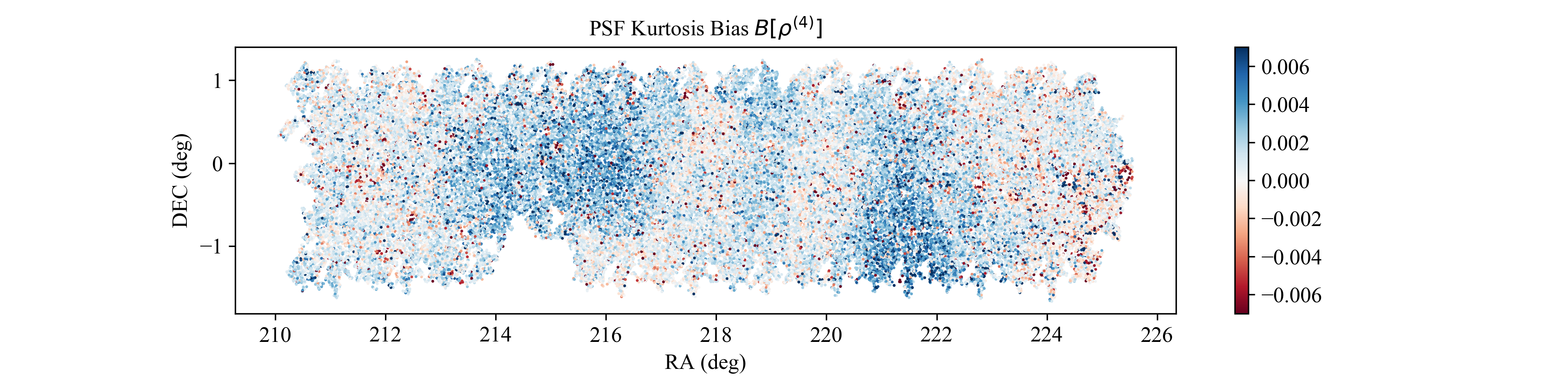}
    \caption{\label{fig:hsc_map}HSC PSF kurtosis as a function of position in the GAMA15H field: true kurtosis (top), and residual of the model $B[\rho^{(4)}]$ (bottom). Each point in the plots represents one star we choose to measure kurtosis. The value for the truth and the model is the weighted radial kurtosis $\rho^{(4)}$,  and the residual is the fractional error $B[\rakurto]$. \response{The results in the true PSF kurtosis contain shot noise in the image. However, in Sec.~\ref{sec:star_selection}, we show that stars with SNR exceeding $10^3$ have $\delta B[\rakurto] < 0.001$, thus our results here are not heavily affected. 
    }
    }
\end{figure*}

So far, we have developed an understanding of the weak lensing shear bias for a given level of PSF kurtosis bias, with increasingly complex galaxy populations.  In this subsection, we now change direction to assess the typical level of PSF kurtosis bias in one ongoing weak lensing survey, using the HSC star catalog described in Sec.~\ref{sec:hscdata}.  Doing so will enable us to assess the resulting level of weak lensing shear bias, and eventually place requirements on PSF model quality for upcoming surveys such as LSST.  Our assessment involves measuring the moments of \response{coadded i-band star images and the i-band PSF model at their positions (see Sec.~\ref{sec:hscdata} for more details)}.

In Fig.~\ref{fig:hsc_map}, we show maps of the true PSF radial kurtosis and the residual kurtosis $B[\rakurto]$ for one of the six fields in the HSC PDR1. 
The total range of variation in the truth and model kurtosis is around 5\%, 
with an average around $2.16$.  
According to Table~\ref{tab:kurtosis}, the HSC PSF typically has a slightly higher radial kurtosis than Kolmogorov and \responsemnras{Moffat} PSF, meaning that it has relatively larger tails. There is some spatial structure in the true PSF kurtosis, which is captured well by the PSF model. The kurtosis bias of the PSF model is typically less than 0.5 per cent of the true PSF kurtosis, and also exhibits spatial structure. The average kurtosis bias is $\langle B[\rakurto] \rangle = 0.0011$. In Sec~\ref{sec:redshift_dependent_bias}, we show that the two point statistics of weak lensing shear are only impacted by the mean multiplicative bias. This mean kurtosis bias is the key result we need in order to estimate the mean multiplicative bias in shear. As an aside to understand this result, we discuss the one- and two-point functions of these PSF model moment residuals in Appendix~\ref{ap:one_two_point_statistics}. \response{In general, the second moment properties of our PSF samples matches what is found in \cite{2018PASJ...70S...5B} and \cite{2018PASJ...70S..25M}.}

\subsection{Redshift Dependent Bias}
\label{sec:redshift_dependent_bias}

Our final step is to synthesize the results from Subsec.~\ref{sec:great3results} and~\ref{sec:realPSF} to estimate the level of redshift-dependent shear bias due to errors in the higher moments of the PSF for HSC-like PSF modeling quality.  According to The LSST Dark Energy Science Collaboration Science Requirements Document version 1 \citep[hereafter referred to as the DESC~SRD;][]{2018arXiv180901669T}, the redshift-dependent multiplicative shear bias should not exceed 0.013 for Y1 or 0.003 for Y10. This requirement is determined based on a quantity $m_0$ defined by a linear parameterization, 
\begin{equation}\label{eq:m0}
    m(z) = m_0 \left(\frac{2z - z_\text{max}}{z_\text{max}}\right) + \bar{m},
\end{equation}
where $\bar{m}$ is a constant multiplicative bias\response{, which is removed beforehand and omitted in the definition of the DESC~SRD}.
Our prediction is based on the image simulation with COSMOS-based S\'ersic profile galaxies and a Kolmogorov PSF, and kurtosis bias measured by HSC PSF. Notice that we 
approximate the impact of the HSC PSF model residuals in simulations with a simple Kolmogorov PSF, which means \response{we expect} some uncertainty on the order of tens of per cent for our prediction in this Section, \response{as our previous results show this is the level of difference between results with a Gaussian versus a Kolmogorov PSF}.  This is an acceptable uncertainty for this initial pathfinder uncertainty to estimate the approximate level of ensemble shear bias from errors in PSF higher moments.

More quantitatively, the multiplicative bias of a tomographic source bin can be estimated by multiplying the average kurtosis bias (Subsec.~\ref{sec:realPSF}) by the ratio of shear bias to PSF kurtosis bias at the redshift of the tomographic bin (Fig.~\ref{fig:derived_quant}).   
From this figure, we can approximate $m_0$ using $\partial m(z)/ \partial B[\rakurto]$ and the average PSF kurtosis bias $\langle B[\rakurto] \rangle$  in HSC. Setting $z_{\text{max}} = 2.4$, \response{$\partial m_0/\partial B[\rakurto] = -0.46 \pm 0.05$. Adopting $\langle B[\rakurto] \rangle = 0.0011$ as in Section~\ref{sec:realPSF}, we get a redshift-dependent multiplicative bias parameter $m_0 = -0.0005 \pm 0.0001$}.  However, the requirements on multiplicative shear bias mentioned above are meant to cover all sources of shear bias\responsecwr{, and there might be multiplicative bias caused by other higher moments as well}.  For consistency with the method of allocating the systematic error budget in the DESC~SRD, since there are many sources of systematic biases when estimating weak lensing shear, we can only allocate $\sim 1/3$ of the above budget to errors in the PSF-related systematics. Therefore, while this is a factor of four below the goal for Y1 results, it is comparable to the error budget for Y10, motivating further improvements in PSF modeling methodology and incorporation of tests of higher moments of the PSF model during the course of the survey. Notice that this prediction is carried out with PSFEx, which is not the planned PSF modeling algorithm for either LSST Y1 or Y10. Therefore, we expect different modeling quality and shear bias in the actual LSST survey. 

\response{Our fitting to the redshift-dependent multiplicative biases is subject to cosmic variance within the COSMOS dataset, as discussed in \cite{2015MNRAS.449.3597K}. This source of uncertainty will not be reflected in bootstrap errorbars or other internal uncertainty estimates, which means that the fit residuals may noticeably exceed the errorbars.  However, this does not affect the conclusions of this paper. } \responsemnras{Furthermore, our results are based on the shear response to the Kolmogorov PSF, and can be different for the other PSF models we explored in Fig.~\ref{fig:GGsize_ratio}. Based on our run of the COSMOS dataset on the Moffat PSF, we observe a shift and a $\sim 50\%$ change to the slope, compare to the redshift-dependent multiplicative biases $\partial m(z)/ \partial B[\rakurto]$ of the Kolmogorov PSF in Fig.~\ref{fig:derived_quant}.  Considering that the shear responses to all kinds of PSFs follow the same trend, we do not expect the redshift-dependent multiplicative biases of them to be significantly different in the order-of-magnitude. }

\response{Finally, we emphasize that the galaxy population we adopt is not a fully realistic realization of what LSST will observe at full survey depth. This introduces some additional uncertainty on our predictions of shear bias due to PSF higher moments error. However, our results illustrate at the order-of-magnitude level that errors in modeling the higher moments of the PSF modeling are a non-negligible source of systematic uncertainty for weak lensing with LSST. }

\section{Conclusion}
\label{sec:conclusion}

In this paper, we carried out an initial exploration of the impact of errors in the higher moments (beyond second moments) of  PSF models on weak lensing shear measurement. We used image simulations with parametric galaxies (at various levels of complexity) and PSF models produced using \textsc{GalSim} to study how errors in the higher moments of the PSF impact galaxy shape measurement and ensemble weak lensing shear measurement. 
We used images of stars and the associated PSF models in the HSC PDR1 data to measure the bias in PSF model kurtosis for real PSFs estimated with \textsc{PSFEx}. Combining the simulation and the HSC results, we found that the current level of errors in the \responsecwr{kurtosis} of the PSF model in HSC can cause \response{$\sim$0.05\%} multiplicative bias in shear measurement. 

There are a number of simplifications associated with our work. In this paper, we only quantified errors in the radial kurtosis of the PSF model. The resulting galaxy shape and lensing shear systematics are purely multiplicative as a result (shear error proportional to input shear) and are also directly proportional to the kurtosis difference between the model and true PSF. We found that the derivative of the linear relationship between shear bias and kurtosis bias depends primarily on the size ratio of the galaxy to the PSF; this relationship  is not monotonic. \responsemnras{We conduct tests of such effect on three different parametric PSFs and a non-parametric PSF, and find similar trends among all results.} Dependencies on the galaxy S\'ersic index and the galaxy shape measurement method are significantly weaker.  \responsecwr{Comparing to} the findings of \cite{2020A&A...636A..78S}, we did not see significant shape measurement dependence for the shear bias induced by errors in the higher moments of the PSF.  Since the PSF higher moment residuals in \cite{2020A&A...636A..78S} are more complicated than our radial moments residual, further simulation on other moments is needed to understand the difference. Furthermore, \cite{2020A&A...636A..78S} studies the PSF of space-based telescope, while we are focusing on ground-based PSF: the difference between the base PSF might also induce a different response from certain shape measurement methods. \responsecwr{Lastly, the shape measurement methods compared in this paper are less different than those compared in \cite{2020A&A...636A..78S}, which could easily explain the different findings. }

We used stars with high signal-to-noise ratio in HSC coadded images, together with the PSF model at the star positions, to measure the errors in the radial weighted kurtosis of the HSC PDR1 PSF models. We found that the kurtosis error of the PSF model is on average $\sim$0.1 per cent, but can be as large as $\sim$1 per cent of the true kurtosis value. The PSF model kurtosis tends to be overestimated for smaller PSFs. 

Finally, we used the COSMOS parametric catalog to simulate the impact of PSF model kurtosis biases on weak lensing shear measurement with a galaxy population that has a realistic distribution of galaxy sizes, shapes, and S\'ersic indices.  Our results suggest that the resulting shear biases are redshift-dependent, primarily due to the shear biases depending on the galaxy apparent size (which is redshift-dependent). The redshift-dependent multiplicative bias $m_0$, defined in Eq.~\eqref{eq:m0}, which can affect cosmological parameter constraints, is estimated as being roughly 0.05\%, at the level of Y10 requirements for LSST. 

While our results show that the ensemble weak lensing shear bias caused by errors in PSF higher moments are not a concern for the current generation of ground surveys, e.g. DES, HSC and KiDS, it is large enough that future surveys such as LSST will need to address this challenge. We see several implications from this study. First, the development of future PSF modeling algorithms should include tests of the fidelity of recovering PSF higher moments, rather than just the second moments.  Second, future surveys should explicitly test the modeling quality of PSF higher moments as part of their science verification process. Finally, this paper also motivates future work on more detailed and rigorous analysis on the shear bias associated with errors in the higher moments of the PSF model. One limitation of the analysis carried out in this pathfinding work is that we are changing multiple moments at a time, while using the radial kurtosis as a single proxy for the impact of higher moments. A more rigorous future analysis requires consideration of individual higher moment and analysis of their impact to weak lensing, as well as a guidelines for placing requirements on the modeling fidelity for these moments.

\section*{Acknowledgments}

This paper has undergone internal review in the LSST Dark Energy Science Collaboration by Arun Kannawadi, Hironao Miyatake, and Pierre Astier, we thank them for their constructive comments and reviews. We thank Mike Jarvis, and Josh Meyers for the helpful comments and discussion. \responsecwr{We thank Morgan Schmitz and Douglas Clowe for the constructive inputs on the paper.} \responsemnras{We thank the anonymous referee for their thoughtful and valuable feedback on the paper.}

TZ and RM are supported in part by the Department of Energy grant DE-SC0010118 and in part by a grant from the Simons Foundation (Simons Investigator in Astrophysics, Award ID 620789). TZ developed the simulation and measurement software, carried out analysis on the results, and led the writing of the manuscript. RM proposed the project, advised on the motivation, experimental design and analysis, and edited the manuscript. 

The DESC acknowledges ongoing support from the Institut National de Physique Nucl\'eaire et de Physique des Particules in France; the Science \& Technology Facilities Council in the United Kingdom; and the Department of Energy, the National Science Foundation, and the LSST Corporation in the United States.  DESC uses resources of the IN2P3 Computing Center (CC-IN2P3--Lyon/Villeurbanne - France) funded by the Centre National de la Recherche Scientifique; the National Energy Research Scientific Computing Center, a DOE Office of Science User Facility supported by the Office of Science of the U.S.\ Department of Energy under Contract No.\ DE-AC02-05CH11231; STFC DiRAC HPC Facilities, funded by UK BIS National E-infrastructure capital grants; and the UK particle physics grid, supported by the GridPP Collaboration.  This work was performed in part under DOE Contract DE-AC02-76SF00515.

Based in part on data collected at the Subaru Telescope and retrieved from the HSC data archive system, which is operated by Subaru Telescope and Astronomy Data Center at National Astronomical Observatory of Japan.

The Hyper Suprime-Cam Subaru Strategic Program (HSC-SSP) is led by the astronomical communities of Japan and Taiwan, and Princeton University.  The instrumentation and software were developed by the National Astronomical Observatory of Japan (NAOJ), the Kavli Institute for the Physics and Mathematics of the Universe (Kavli IPMU), the University of Tokyo, the High Energy Accelerator Research Organization (KEK), the Academia Sinica Institute for Astronomy and Astrophysics in Taiwan (ASIAA), and Princeton University.  The survey was made possible by funding contributed by the Ministry of Education, Culture, Sports, Science and Technology (MEXT), the Japan Society for the Promotion of Science (JSPS),  (Japan Science and Technology Agency (JST),  the Toray Science Foundation, NAOJ, Kavli IPMU, KEK, ASIAA, and Princeton University.

This paper makes use of software developed for the Large Synoptic Survey Telescope. We thank the LSST Project for making their code available as free software at http://dm.lsst.org. 

The Pan-STARRS1 Surveys (PS1) have been made possible through contributions of the Institute for Astronomy, the University of Hawaii, the Pan-STARRS Project Office, the Max-Planck Society and its participating institutes, the Max Planck Institute for Astronomy, Heidelberg and the Max Planck Institute for Extraterrestrial Physics, Garching, The Johns Hopkins University, Durham University, the University of Edinburgh, Queen’s University Belfast, the Harvard-Smithsonian Center for Astrophysics, the Las Cumbres Observatory Global Telescope Network Incorporated, the National Central University of Taiwan, the Space Telescope Science Institute, the National Aeronautics and Space Administration under Grant No. NNX08AR22G issued through the Planetary Science Division of the NASA Science Mission Directorate, the National Science Foundation under Grant No. AST-1238877, the University of Maryland, and Eotvos Lorand University (ELTE) and the Los Alamos National Laboratory.

We thank the developers of \textsc{GalSim}, \textsc{ngmix}, and \textsc{TreeCorr} for making their software openly accessible. 

\section*{Data Availability}

The HSC-SSP data in this paper is publicly available at \url{https://hsc-release.mtk.nao.ac.jp/doc/index.php/tools-2/}. 
The COSMOS catalog is available at \url{https://zenodo.org/record/3242143#.YF2bHK9KiUk}. Simulation and analysis code will be shared on reasonable request to the authors.

\bibliographystyle{mnras}
\bibliography{main}

\appendix

\section{Statistics of the HSC PSF}
\label{ap:one_two_point_statistics}

\begin{figure}
    \centering
    \includegraphics[width=0.8\columnwidth]{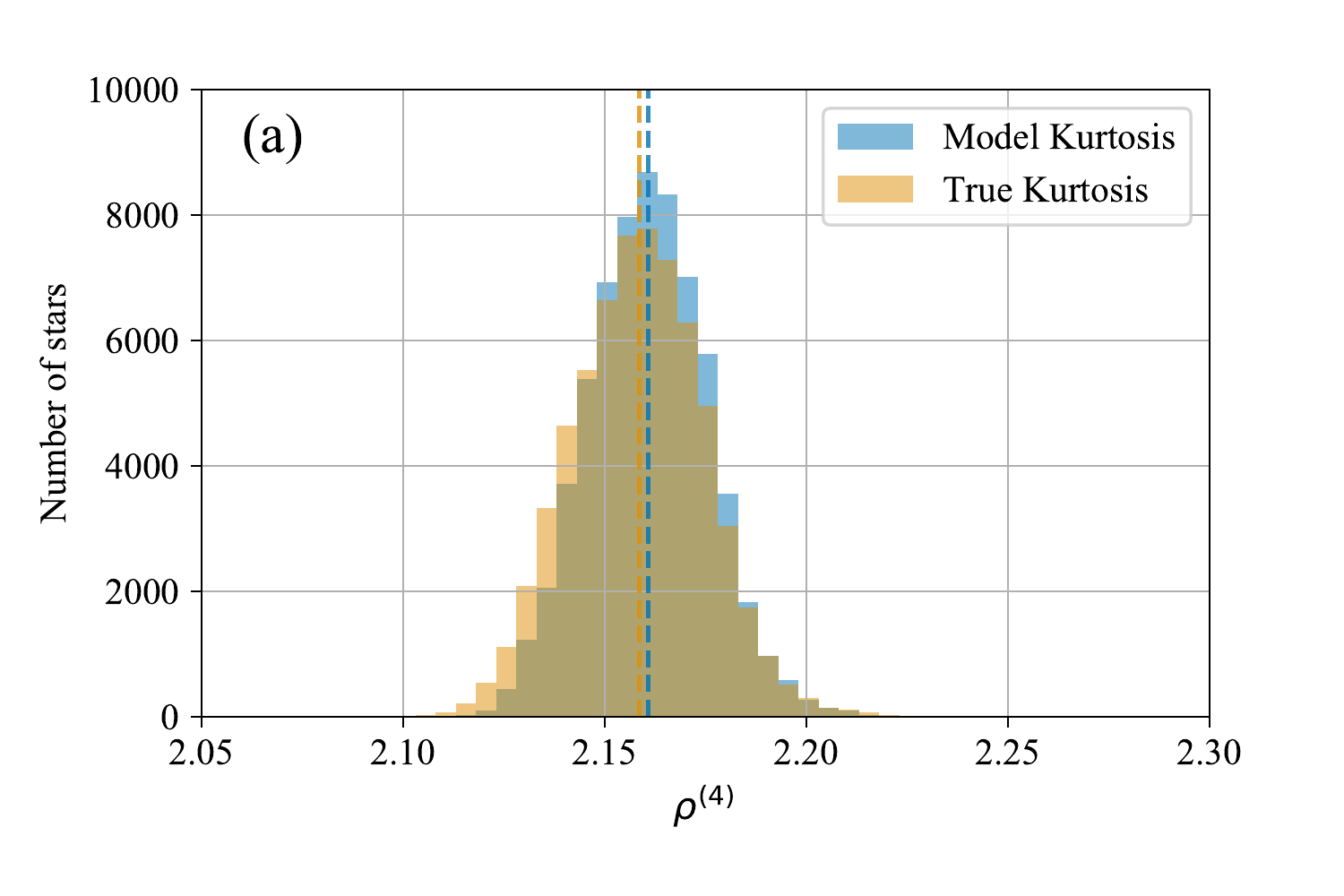}
    \includegraphics[width=0.8\columnwidth]{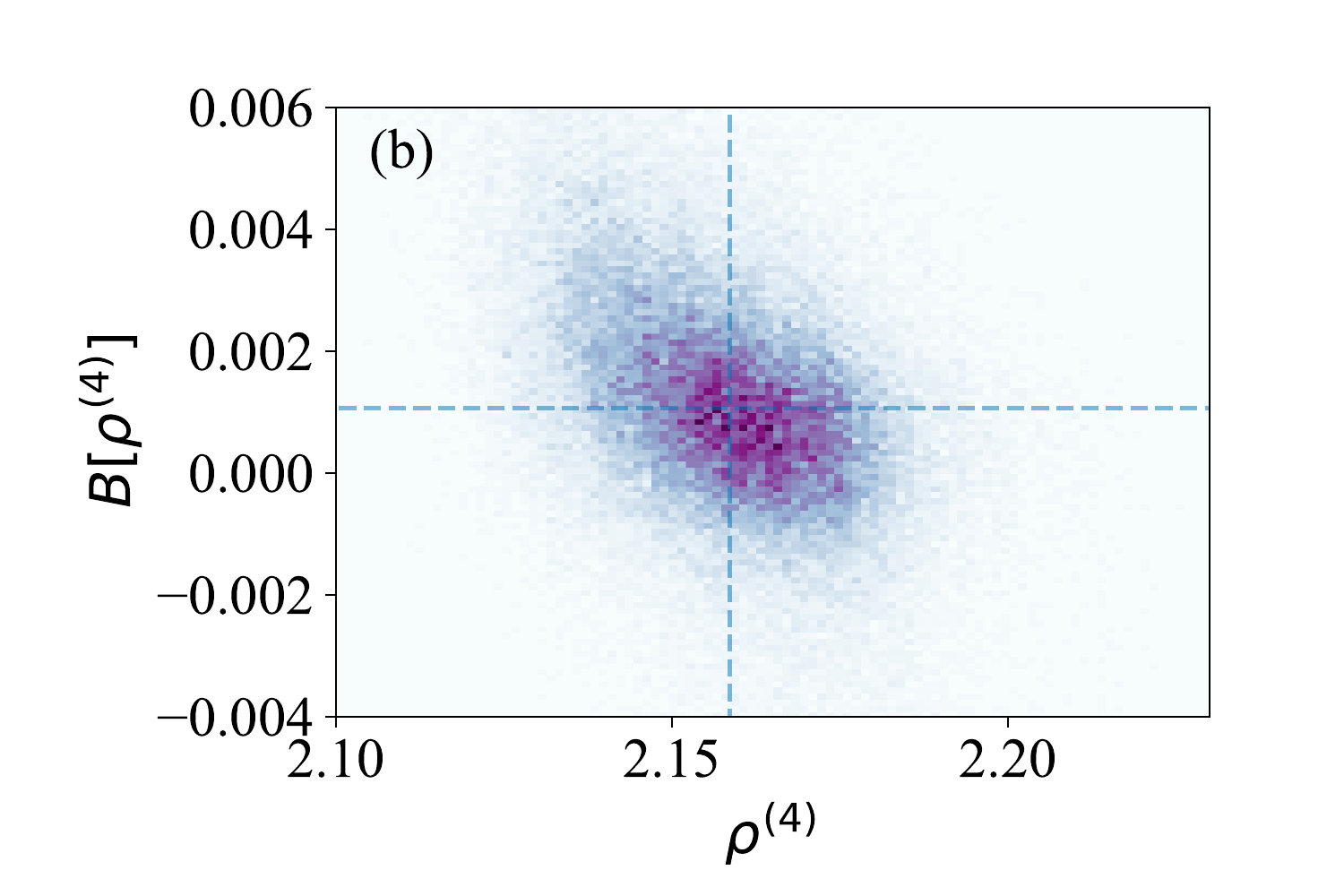}
    \includegraphics[width=0.8\columnwidth]{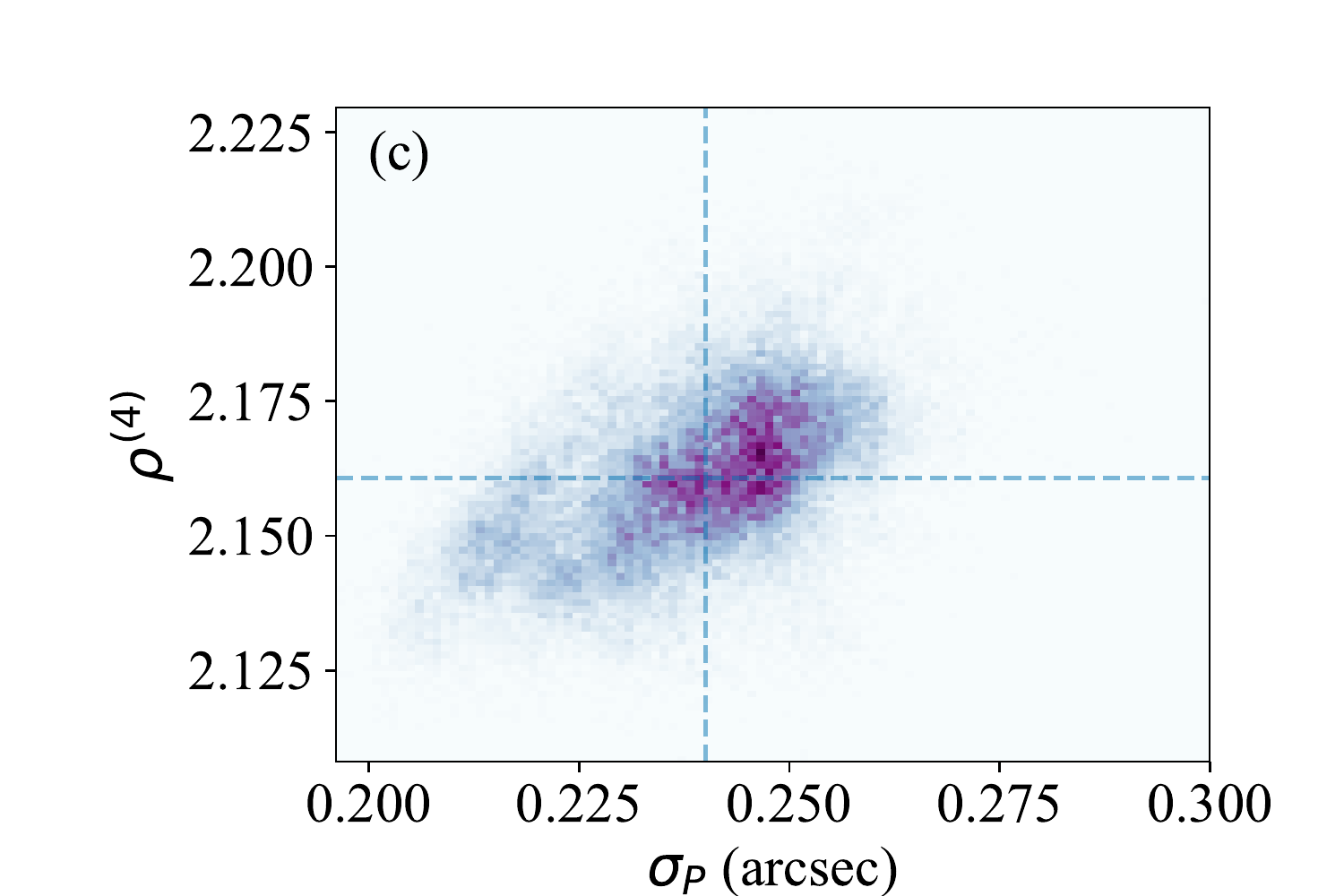}
    \includegraphics[width=0.8\columnwidth]{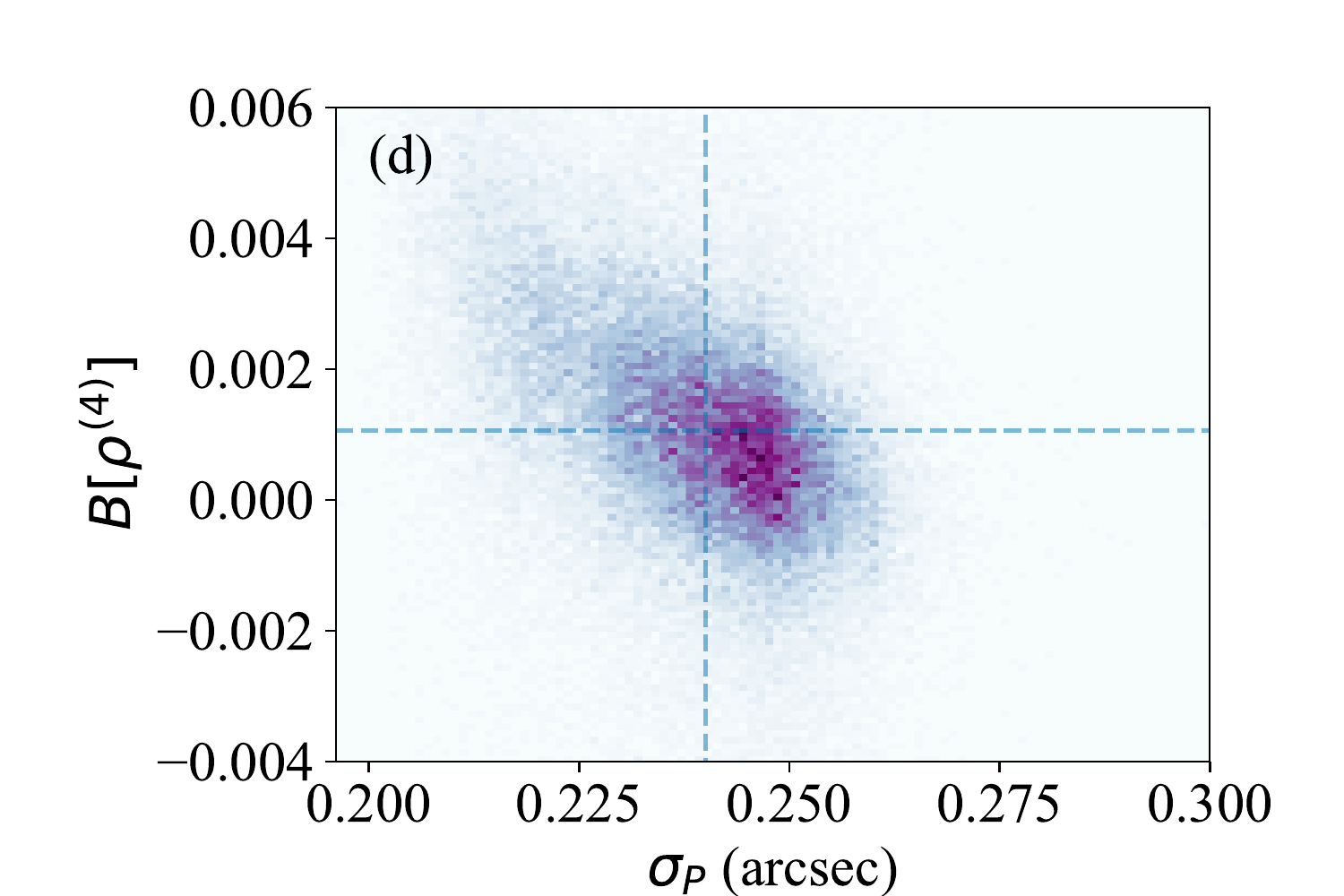}
    \caption{ \textbf{(a)}: 1D distributions of the true and model PSF kurtosis. 
    \textbf{(b)}: 2D distribution of the kurtosis bias $B[\rakurto]$ and PSF size $\sigma_P$; \textbf{(c)}: 2D distribution of the true kurtosis $\rakurto$ and the PSF size $\sigma_P$; \textbf{(d)}: 2D distribution of the kurtosis bias $B[\rakurto]$ and the PSF size $\sigma_P$. For the three 2D distribution plots, the median values of the quantities on each axis are shown with dashed lines. The color scales of the distributions are linear in the density. 
    }
    \label{fig:hsc_1pt}
\end{figure}

In Fig.~\ref{fig:hsc_1pt}, we show the 1D histogram of PSF true $\rakurto$ and model kurtosis $\hat{\rho}^{\text{(4)}}$ , and 2D histogram of the true kurtosis $\rakurto$, its bias $B[\rakurto]$, and their connection to the PSF size. 
0.1\% larger than the true PSF kurtosis in panel~\textbf{(a)}. The width of the true PSF kurtosis distribution is slightly larger than the width of the model PSF kurtosis.   A smaller PSF generally has a smaller kurtosis. This is likely caused by a more substantial contribution from the Airy PSF, for smaller PSFs, as Table~\ref{tab:kurtosis} suggests that Airy PSFs have a smaller kurtosis than Kolmogorov PSFs.   
Also, when the PSF is small, the PSF model tends to overestimate the kurtosis. 
\response{Note that there is shot noise in the images we measure. However, its impact on this figure is negligible, since our simulation in Sec.~\ref{sec:star_selection} shows that the uncertainty in $B[\rakurto]$ due to shot noise is $<0.001$, and the effect on the average $B[\rakurto]$ is $<0.001/\sqrt{n}$, orders of magnitude smaller than our average bias measured.}

\end{document}